\documentclass[reprint, amsmath,amssymb,aps,twocolumn, prb,superscriptaddress]{revtex4-1}
\usepackage{amsfonts,amssymb}
\usepackage{mathrsfs}
\usepackage{amsmath}
\usepackage[none]{hyphenat}
\usepackage{graphicx}
\usepackage{hyperref}    
\usepackage{bm}          
\usepackage{natbib}
\usepackage{soul} 
\usepackage{color}
\usepackage{longtable}
\usepackage{ textcomp }
\usepackage{verbatim}

\begin{document}

\title{Hopping conductivity and insulator-metal transition in films of touching semiconductor nanocrystals}

\author{Han Fu}
\email{fuxxx254@umn.edu}
\affiliation{Fine Theoretical Physics Institute, University of Minnesota, Minneapolis, MN 55455, USA}
\author{K. V. Reich}
\affiliation{Fine Theoretical Physics Institute, University of Minnesota, Minneapolis, MN 55455, USA}
\affiliation{Ioffe Institute, St Petersburg, 194021, Russia}
\author{B. I. Shklovskii}
\affiliation{Fine Theoretical Physics Institute, University of Minnesota, Minneapolis, MN 55455, USA}
\date{\today}

\begin{abstract}
This paper is focused on the the variable-range hopping of electrons in semiconductor nanocrystal (NC) films below the critical doping concentration $n_c$ at which it becomes metallic. The hopping conductivity is described by the Efros-Shklovskii law which depends on the localization length of electrons. We study how the localization length grows with the doping concentration $n$ in the film of touching NCs. For that we calculate the electron transfer matrix element $t(n)$ between neighboring NCs for two models when NCs touch by small facets or just one point. We study two sources of disorder: variations of NC diameters and random Coulomb potentials originating from random numbers of donors in NCs. We use the ratio of $t(n)$ to the disorder-induced NC level dispersion to find the localization length of electrons due to the multi-step elastic co-tunneling process. We found three different phases at $n<n_c$ depending on the strength of disorder, the material, sizes of NCs and their facets: 1) ``insulator'' where the localization length of electrons increases monotonically with $n$ and 2) ``oscillating insulator'' when the localization length (and the conductivity) oscillates with $n$ from the insulator base and 3) ``blinking metal'' where the localization length periodically diverges. The first two phases were seen experimentally and we discuss how one can see the more exotic third one. In all three the localization length diverges at $n=n_c$. This allows us to find $n_c$.
\end{abstract}

\maketitle

\section{Introduction}

Semiconductor nanocrystals (NCs) have a great potential for optoelectronics applications such as solar cells \cite{Geier}, light emitting diodes \cite{Wood} and field effect transistors \cite{Kikkawa_2014,gresback_controlled_2014}. Their advantage is size-tunable optical and electrical properties \cite{Alivisatos_1996} and low-cost solution-based processing techniques \cite{Sargent,Murray}. These applications require conducting NC films and several ways of introducing carriers via doping are being explored \cite{Wang_2001,Yu_2003,Uttrecht,Liu_2010,Banin_2011,Norris_2012,HgS,Kikkawa_2014,Ting}. At a given concentration of carriers one tries to improve the mobility by moving NCs closer to each other and reducing their contact resistance.

In many studies~\cite{Wang_2001,Yu_2003,Ting,Liu_2010} the low temperature conductivity of doped films was found to obey Efros-Shklovskii (ES) variable range hopping law~\cite{ES_1975}:
\begin{equation}
\label{eq:ES}
\sigma(T) = \sigma_0 \exp\left[-\left(\frac{T_{ES}}{T}\right)^{1/2}\right].
\end{equation}
Here $\sigma_0$ is a conductivity prefactor, $T$ is the temperature, and
\begin{equation}
\label{eq:TES}
T_{ES}=\frac{Ce^2}{\varepsilon_f k_B \xi},
\end{equation}
in Gaussian units.
Here $e$ is the electron charge, $\xi$ is the localization length, $\varepsilon_f$ is the effective dielectric constant of the film, $k_B$ is the Boltzmann constant, $C\simeq9.6$ \cite{Boris}. Typically, $\xi$ grows with the concentration of electrons $n$ in a NC and with the improvement of contacts between NCs. Therefore, $T_{ES}$ becomes smaller and the film becomes more conducting~\cite{Ting}.

\begin{figure}[h]
$\begin{array}{c}
\includegraphics[width=0.5\textwidth]{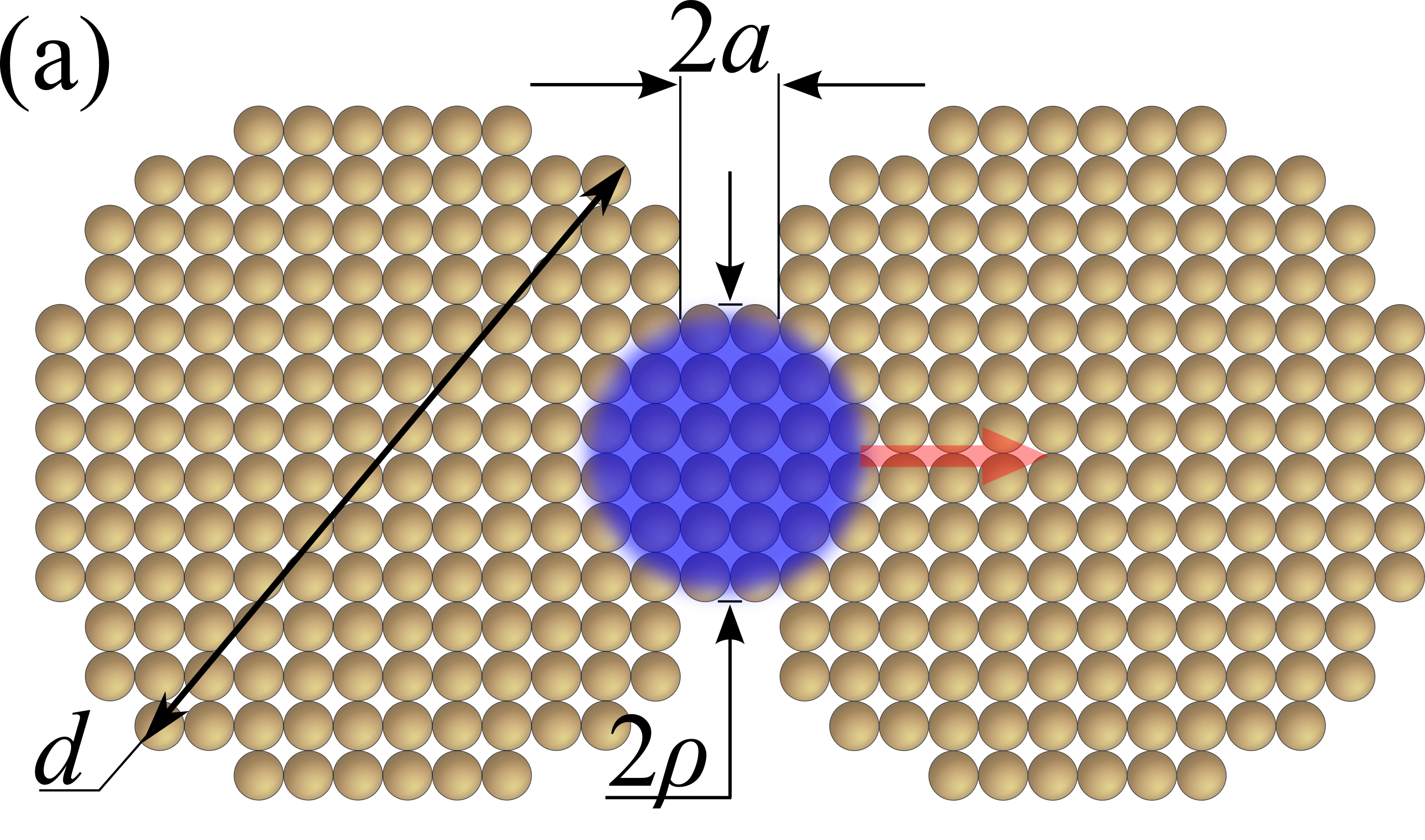}\\
\includegraphics[width=0.5\textwidth]{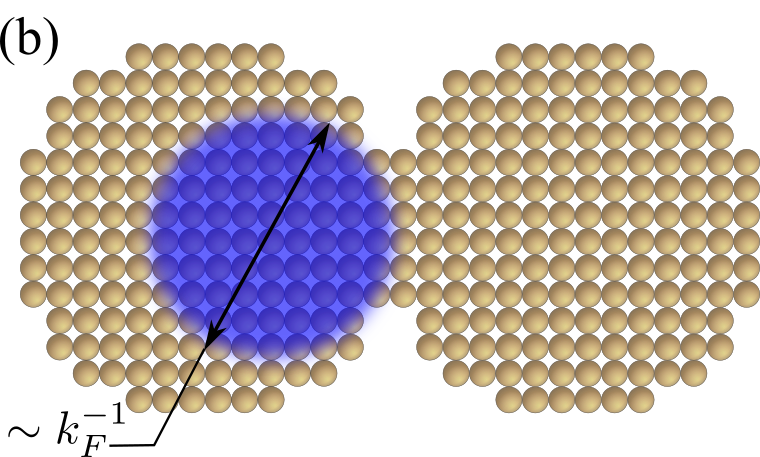}
\end{array}$
\caption{(Color online) The cross-section of two NCs in contact by their facets with radius $\rho\ll d$ each. Here $a$ is the lattice constant, $d$ is the NC diameter. The blue cloud depicts the smallest available electron wave packet with the size $k_F^{-1}\sim n^{-1/3}$, where $k_F$ is the Fermi wavenumber and $n$ is the doping concentration of electrons in each NC. (a) Electron transport at $n > n_c$. The smallest electron packet fits in the touching facets and moves through the contact. (b) At $n < n_c$, the smallest wave packet gets stuck near the contact and the electron tunneling between NCs is depleted so much that it cannot overcome the disorder to delocalize electrons.}\label{fig:0}
\end{figure}

In this paper we concentrate on doping of NC films by chemical donors or acceptors \cite{Norris_2008} which was recently achieved in InAs~\cite{Banin_2011}, CdSe \cite{Norris_2012}, HgS \cite{HgS} and Si~\cite{Ting} NCs. While many experimental studies have been directed towards increasing the conductivity of NC films with increased $n$, it was not clear when $\xi$ diverges and $T_{ES}$ vanishes so that the NC film becomes metallic \cite{guyot-sionnest_electrical_2012, shabaev_dark_2013, Band_like_transport_review}. In other words, what is the critical concentration $n_c$ of electrons (or donors) in a NC necessary for the insulator-metal transition (IMT)? Recently~\cite{Ting} $n_c$ was estimated for the case favorable for the IMT , where close-to-spherical NCs touch each other by small facets of radius $\rho\ll d$ without any ligands that impede the conduction by creating a barrier between NCs (see Fig. \ref{fig:0}a). The result is very simple
\begin{equation}
\label{eq:result}
n_c \simeq 0.3 \rho^{-3}.
\end{equation}
The IMT is illustrated in Fig. \ref{fig:0}a where we show how an electron wave packet of the minimum available size for a given $n$ quasiclassically passes between two touching NCs at $n > n_c$, but has to tunnel at $n < n_c$ and, therefore, becomes more vulnerable to disorder.

Contacts between NCs may have different origins. For example, a close-to-spherical NC has small facets due to the discreteness of the crystal lattice. Their radius can be estimated as $\rho_a=\sqrt{da/2}$, where $a$ is the lattice constant and $d$ is the NC diameter. For CdSe NCs with $a=0.6$ nm and $d=5$ nm, $\rho_a \sim 1.2$ nm and Eq. (\ref{eq:result}) gives $n_c = 2 \times 10^{20}\, \mathrm{cm}^{-3}$. For the case in which NCs shown in Fig. \ref{fig:0}a touch each other away from these facets, a finite tunneling distance $b \sim 0.1$ nm in the medium between NCs should be taken into account. This leads to Eq. (\ref{eq:result})
where $\rho=\rho_b=\sqrt{db/2}\ll \rho_a$ is the radius of an effective ``$b$-contact" and the critical concentration $n_c$ is much larger.

On the other hand, at very light doping when the average number of electrons per NC, $N = \pi n d^3/6$, is less than unity one should see the nearest-neighbor hopping between NCs with the activation energy equal to the charging energy of a NC~\cite{Boris,Ting_2014}. Thus, the ES hopping should be observed in a large range of the concentrations $1/d^{3} < n < n_c$. To calculate $T_{ES}$ given by Eq. \eqref{eq:TES}, we need to know how the localization length $\xi(n)$ grows in this range, before reaching the NC diameter $d$ and diverging in a critical vicinity of $n_c$.

The localization length of electrons is determined by the competition of the disorder energy $\delta E$ and the tunneling matrix element $t$ between neighboring NCs. We study two main sources of disorder: the dispersion of NC diameters, which changes the quantization kinetic energy, and the variation of the number of donors in a NC, which leads to charging of NCs and random Coulomb potentials shifting electron levels. We also calculate $t(n,\rho)$ for two mentioned above models of small-$\rho$ contacts. We arrive to the conclusion that typically the combination of both sources of disorder is so strong that one needs $N \gg 1$ electrons per NC to make large enough $t$ in order to get appreciable $\xi$ and approach the IMT.

In this paper we deal with the generic case for small semiconductor NCs when electron energy shells of the spherical NCs are weakly split and separated by the quantization gap $\Delta$ . We show that when the disorder energy $\gamma$ becomes larger than $\Delta$ and NCs touch by contact facets of small radius $\rho$, the localization length $\xi$ is
\begin{equation}
\xi(n)\approx\frac{d}{\ln(2/n\rho^3)}.\label{eq:4}
\end{equation}
This result is obtained at low temperatures when electrons hop via the elastic multi-step co-tunneling between distant NCs.  The effective dielectric constant $\varepsilon_f$ is not affected by the electron polarization in NCs far from the IMT. Thus, Eq. (\ref{eq:4}) together with Eq. \eqref{eq:TES} can be used to predict the dependence $T_{ES} (n)$.
The localization length $\xi$ becomes comparable to $d$ when $n$ gets close to the critical concentration $n_c$ of the IMT. Using this criterion, we arrive from the insulating side at the estimate Eq. (\ref{eq:result}) which was obtained in Ref. \onlinecite{Ting} from the metallic side.

Both Eq. \eqref{eq:4} and Eq. \eqref{eq:result} do not depend on the disorder strength. This happens because when $\gamma$ exceeds $\Delta$ the energy difference between neighboring NCs $\delta E$ saturates at $\Delta$ due to the periodicity of the quantized spectrum (see Fig. \ref{fig:dis}). Remarkably both Eq. \eqref{eq:4} and Eq. \eqref{eq:result} continue to play an important role when $\gamma$ becomes smaller than $\Delta$ and well defined peaks of the density of states appear (see Fig. \ref{fig:cotun1}). With decreasing width of these peaks the localization length starts to oscillate at small $N$ while keeping its minima close to the base line Eq. \eqref{eq:4} (see Fig. \ref{fig:oscil}, we call this phase oscillating insulator (OI)). By further decreasing the disorder the oscillations can take over all range of concentrations $n < n_c$ and eventually $\xi$ diverges at series of maximum points adjacent to $n_c$ (see Fig. \ref{fig:osci2}, we call this phase blinking metal (BM)). New phases OI and BM are shown together with the large-disorder phase called here the usual insulator (I) where the localization length obeys Eq. \eqref{eq:4} and the metallic phase (M) in two phase diagrams in the plane $(N, \Delta/E_c)$ (see Figs. \ref{fig:lowN} and \ref{fig:final2}). Note that the border of M is always given by Eq. (\ref{eq:result}).

We used our phase diagrams to address the situation in several widely used semiconductor NCs, i.e., CdSe, InAs and ZnO, with $d=5$ nm, $\rho=\rho_a=1.2$ nm and 7\% dispersion of NC diameters. We show that in CdSe and InAs the Coulomb interaction can ignored (marginally in CdSe) and with the growing $N$ one can see only two phases OI and M. A substantially broad BM phase which, of course, would improve the NC film conductivity at smaller $N$ requires even smaller dispersion of diameters, say 3\%. However, for $\rho\leq \rho_a$ the highly desirable metallic state for $N \sim 1$ when only the 1S shell is filled can be achieved only with unrealistic less-than-1\% dispersion of NC diameters. Of course, one can always increase $\rho$ to achieve BM and extend it all the way till $N=1$ (on the way to the bulk semiconductor). In ZnO Coulomb disorder effects play an important role leading to the expansion of the I phase between OI and M ones.

The paper is organized as follows. In Sec. \ref{sec:sing}, we dwell upon the main energies of a single NC, i.e., the quantization energy gap $\Delta$ separating consecutive degenerate shells of the electron spectrum and the charging energy $E_c$ of a NC. In Sec. \ref{sec:NCd} we start from very large ratios of $\Delta/E_c $ where the dispersion of NC diameters dominates over the Coulomb disorder. We use values of $t(n,\rho)$ calculated later in Sec. \ref{sec:facet} to find $\xi(n)$ and $n_c$. For the case of relatively large diameter dispersion and very small $\rho$ at large $n$, the localization length $\xi(n)$ follows Eq. (\ref{eq:4}) and $n_c$ is given by Eq. (\ref{eq:result}). We also study the case of a very weak diameter dispersion and arrive at BM. In Sec. \ref{sec:charge} we study the charging of NCs and the resulting Coulomb disorder and get $\xi(n)$ for any $\Delta/E_c$. We show that at $\Delta/E_c < 5$ the Coulomb disorder eliminates the BM phase and extends the range of validity of Eq. (\ref{eq:4}). In Sec. \ref{sec:discussion} we discuss examples of widely used semiconductor CdSe, InAs and ZnO NCs. In Sec.~\ref{sec:facet} we calculate the tunneling matrix element $t(n,\rho)$ for NCs touching by contact facets (see Fig. \ref{fig:0}b). In Sec. \ref{sec:bcon}, we study the case when NCs touch each other away from prominent facets or are separated by short ligands and derive the corresponding expressions for $\xi$. In Sec. \ref{sec:nond} we deal with large NCs where the random electric field of donors split and mix degenerate shell levels so that semiconductor NCs acquire random spectra similar to that of metallic granules. We conclude in Sec. \ref{sec:con}.

\section{NC electronic spectrum and charging energy}
\label{sec:sing}

We assume that close-to-spherical NCs have diameter $d$ and touch each other by facets with radius $\rho$. At small enough $\rho$ electrons are localized inside NCs. We  suppose that the electron wave function is close to zero at the NC surface, due to the large confining potential  barrier created by the insulator matrix surrounding the NC. Under these conditions, electrons occupy states with different radial and angular momentum quantum numbers, i.e., $(n,l)$-shells, each being degenerate with respect to the azimuthal quantum number $m=-l,\dots,l$ where the polar axis ($z$ axis) is defined in the direction of electron tunneling connecting centers of two neighboring NCs (we talk more about this in Sec. \ref{sec:facet}). As we explained in Introduction we are interested in NCs with the average electron number $N \gg 1$. Therefore several $(n,l)$-shells are occupied. The quantum energy gap between two consecutive $(n,l)$-shells typically is
\begin{equation}
\label{eq:x}
\Delta \simeq \frac{20\hbar^2}{m^*d^2}
\end{equation}
where $m^*$ is the effective electron mass inside NCs.

Also, when the quantum numbers are large, Bohr's correspondence principle allows us to consider quasiclassically the average density of states of electrons and introduce the Fermi wave number $k_F$
\begin{equation}
\label{eq:Fermi_wave}
k_{F}=\left(3 \pi^2\right)^{1/3} n^{1/3}.
\end{equation}
Here $n=6N/\pi d^3$ is the density of electrons in a NC. Below, $k_F$ serves as a measure of the concentration $n$.

The kinetic energy of electrons is only a part of the total energy of the NC. One should add to it the total Coulomb interaction energy of all electrons and donors. In general,  calculating the total Coulomb energy (self-energy) of the system is a difficult problem because of the random position of donors.  For our case, however, a significant simplification is available because the semiconductor dielectric constant $\varepsilon$ is typically much larger than the dielectric constant $\varepsilon_m$ of the medium in which the NC is embedded. This allows us to ignore in the first approximation the energy dependence on positions of donors and electrons and instead concentrate on only the dependence on the total charge $Qe$ of the NC.

The energy of a NC with charge $Qe$ surrounded by neutral NCs (the self-energy) is equal to $Q^2 E_c$, where the charging energy is
\begin{equation}
  \label{eq:charging_energy}
  E_c=\frac{e^2}{\varepsilon_f d}\,\,.
\end{equation}
For non-touching NCs where the volume fraction of semiconductor NCs is $f\le0.52$, one can use the Maxwell-Garnet formula \cite{Maxwell}
\begin{equation}
\varepsilon_f = \varepsilon_m\frac{\varepsilon+2\varepsilon_m+2f(\varepsilon-\varepsilon_m)}{\varepsilon+2\varepsilon_m-f(\varepsilon-\varepsilon_m)}
\label{eq:dielectric_constant}
\end{equation}
to calculate the effective dielectric constant $\varepsilon_f$. This gives $\varepsilon_f\simeq 3$ at $f=0.52$ corresponding to the very moment of NC touching (we take $\varepsilon_m=1,\,\varepsilon=10$ as in the case of CdSe NCs). For these $\varepsilon_m$ and $\varepsilon$, the effective dielectric constant $\varepsilon_f$ was calculated numerically for all range of $f$ \cite{Liu} including $f>0.52$ obtained for faceted NCs touching by facets. One can check~\cite{Reich} that Eq. \eqref{eq:dielectric_constant} works well even for $f$ as large as $0.7$. This means that for NCs touching by small facets or separated by short ligands, $\varepsilon/\varepsilon_f\simeq 3$ is a good estimate for CdSe and other semiconductors with $\varepsilon\lesssim15$ that we are dealing with in this paper. For semiconductors with much larger $\varepsilon$ one may use results of Ref. \onlinecite{Reich}.

The ratio $\Delta/E_c$ is an important parameter of our theory.
In $n$-type semiconductors we address here, for NCs with $d=5$ nm,  $\Delta/E_c=$2, 3, 5, 27  for Si, ZnO, CdSe, and InAs, respectively.

\section{Localization length and IMT determined by dispersion of NC diameters}
\label{sec:NCd}
There are two important sources of disorder for electrons in a NC film. The first one is the variation of NC diameters. Since the energy gap $\Delta \propto 1/d^2$, each energy level gets a shift $\alpha\Delta$, where $\alpha=2\delta d /d$ and $\delta d$ is the variation of the diameter $d$ (experimentally, $\delta d/d$ is as large as $5-15\%$ \cite{Murray_Monodisperse,Ting} so $\alpha=0.1-0.3$). The second source of disorder is the fluctuations of the donor number in a NC, which result in the charging of NCs and subsequent random potentials. We study this phenomenon in Sec. \ref{sec:charge}. In this section we deal with the case of large enough $\Delta/E_c$ when charging can be ignored.

\begin{figure}[h]
\includegraphics[width=.47\textwidth]{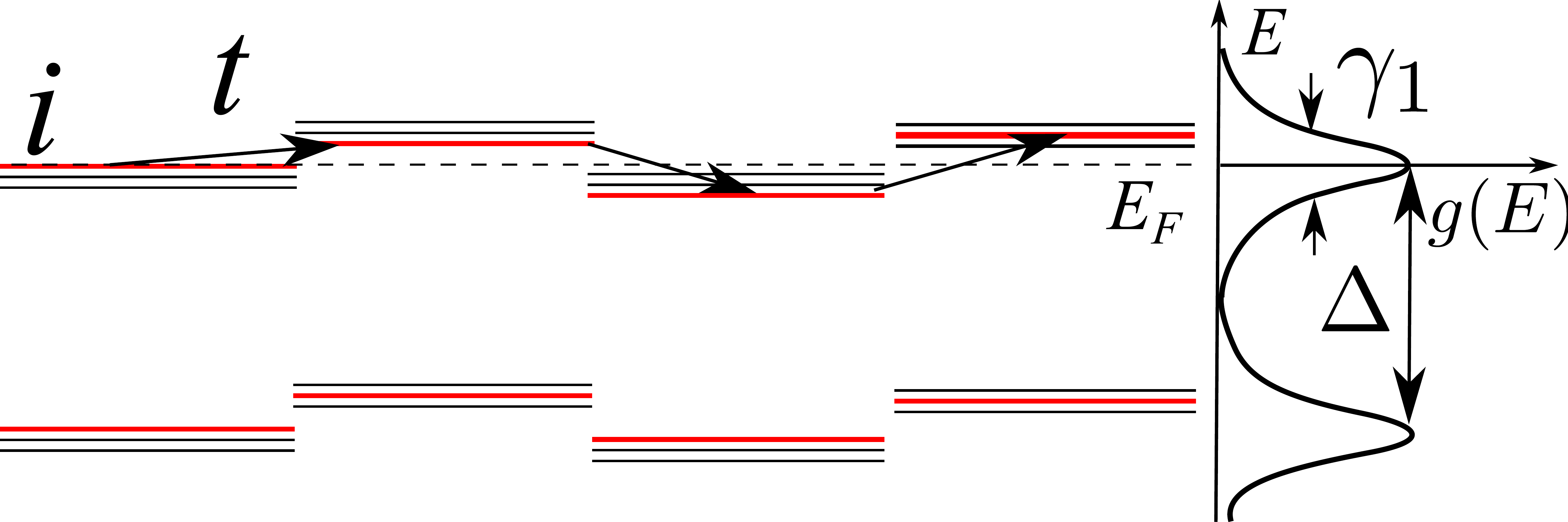}\\
\caption{Energy spectrum of the linear chain of touching spherical NCs at small $N$ in the vicinity of the Fermi level. The degenerate electron levels are aligned with small energy shifts created by the NC diameter variations. The density of states $g(E)$ as a function of the energy $E$ has periodic maxima separated by the energy gap $\Delta$. Peaks of the density of states have width $\gamma_1$. The electron tunnels with the tunneling matrix element $t$ from the initial NC $i$ through the $m=0$ levels (red) in the shell closest to the Fermi level in each intermediate NC. Virtually visited levels are shown by arrows. The dashed line represents the Fermi level.}\label{fig:cotun1}
\end{figure}

The dispersion of NC diameters creates the energy shift of the electron levels close to the Fermi level
\begin{equation}
\gamma_1=N^{2/3}\alpha\Delta,\label{eq:y}
\end{equation}
where $N^{2/3}$ gives the number of filled shells. When $N$ is small, $\gamma_1\ll\Delta$. The energy levels of NCs are then quite aligned and the density of states has periodically alternating maxima and minima (see Fig. \ref{fig:cotun1}). In this case, the transport mechanism depends on the position of the Fermi level or in other words on the average electron number $N$ \cite{Liu_2010, Boris, Boris_2004}. When the Fermi level is in the middle of each degenerate shell, i.e., the local density of states is very large, Coulomb correlations ``dig" the Coulomb gap in this density of states \cite{ES_1975}, which in turn leads to low-temperature ES conductivity law Eq. \eqref{eq:ES}. When the Fermi level is close to the middle of the gap $\Delta$ where a small density of states may be present due to overlapping tails of neighboring shells, the Coulomb effects are not important since the density of states is already very small. Such a constant density of states leads to the Mott variable range hopping \cite{Liu_2010,Boris_2004}.

However, for both ES and Mott variable range hopping, one should use the concept of the localization length which determines the exponential decay of the electron wave function with the distance $x$ from the NC where the localized electron resides in. The localization length $\xi$ is determined by the
co-tunneling between two distant NCs with energies close to the Fermi level \cite{Ioselevich_2005,Beloborodov_2005,Boris_2004,Beloborodov}. In the co-tunneling process, an electron tunnels between neighboring NCs of the chain of $M$ intermediate NCs connecting the initial and final NCs. If after the tunneling all intermediate NCs remain in the ground state, the co-tunneling process is called elastic. Alternatively, an intermediate NC can acquire an electron-hole excitation. Such process is called inelastic. At low temperatures the elastic process dominates.

\begin{figure}[h]
\includegraphics[width=.45\textwidth]{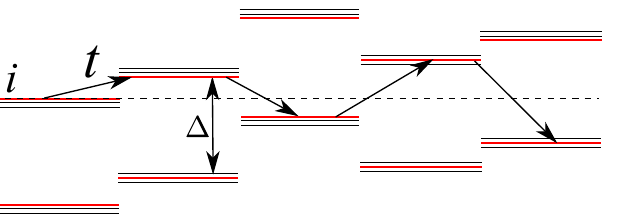}\\
\caption{(Color online) Energy spectrum of a chain of NCs at large average electron number $N$. An electron tunnels from an initial NC $i$ through intermediate NCs. Virtually visited levels are shown by arrows. The dashed line shows the energy of the tunneling electron close to the Fermi level. Each NC has a ladder of $(2l + 1)$-degenerate $(n,l)$-shells with the gap $\Delta$ between them. Due to variations of diameters, the whole ladder of energy levels is shifted up and down by an energy larger than $\Delta$. Here we show only two shells closest to the Fermi level. Only one level (red) of each shell contributes to the tunneling with the matrix element $t$.}\label{fig:dis}
\end{figure}

We show in Sec. \ref{sec:facet} that in the chain of NCs extended along the $z$ direction, inside each intermediate NC only the $m=0$ state in the highly degenerate $(n,l)$-shell contributes to the tunneling process with a dominant matrix element $t$. Thus, along the chain of co-tunneling, there is only one possible series of intermediate energy states closest to the energy of the tunneling electron and no summation over different states of a given shell is needed for calculating the total amplitude.  We can say that we deal with non-degenerate levels (red as shown in Figs. \ref{fig:cotun1} and \ref{fig:dis}) with the energy spacing $\Delta$. This allows us to write estimates for the tunneling amplitude as $\propto \left(t/\delta E\right)^M\simeq e^{-x/\xi}$ where $M=x/d$ is the number of intermediate NCs in the tunneling path and $\xi$ is the localization length
\begin{equation}
\xi\approx\frac{d}{\ln(\delta E/t)}.\label{eq:5}
\end{equation}
Here $\delta E$ is the energy difference between the tunneling electron and the state in the intermediate NC. Eq. \eqref{eq:5} is valid when $\ln(\delta E/t)>1$ or $\xi< d$ and the film is far from the critical vicinity of the IMT.

\begin{figure}[h]
\includegraphics[width=.43\textwidth]{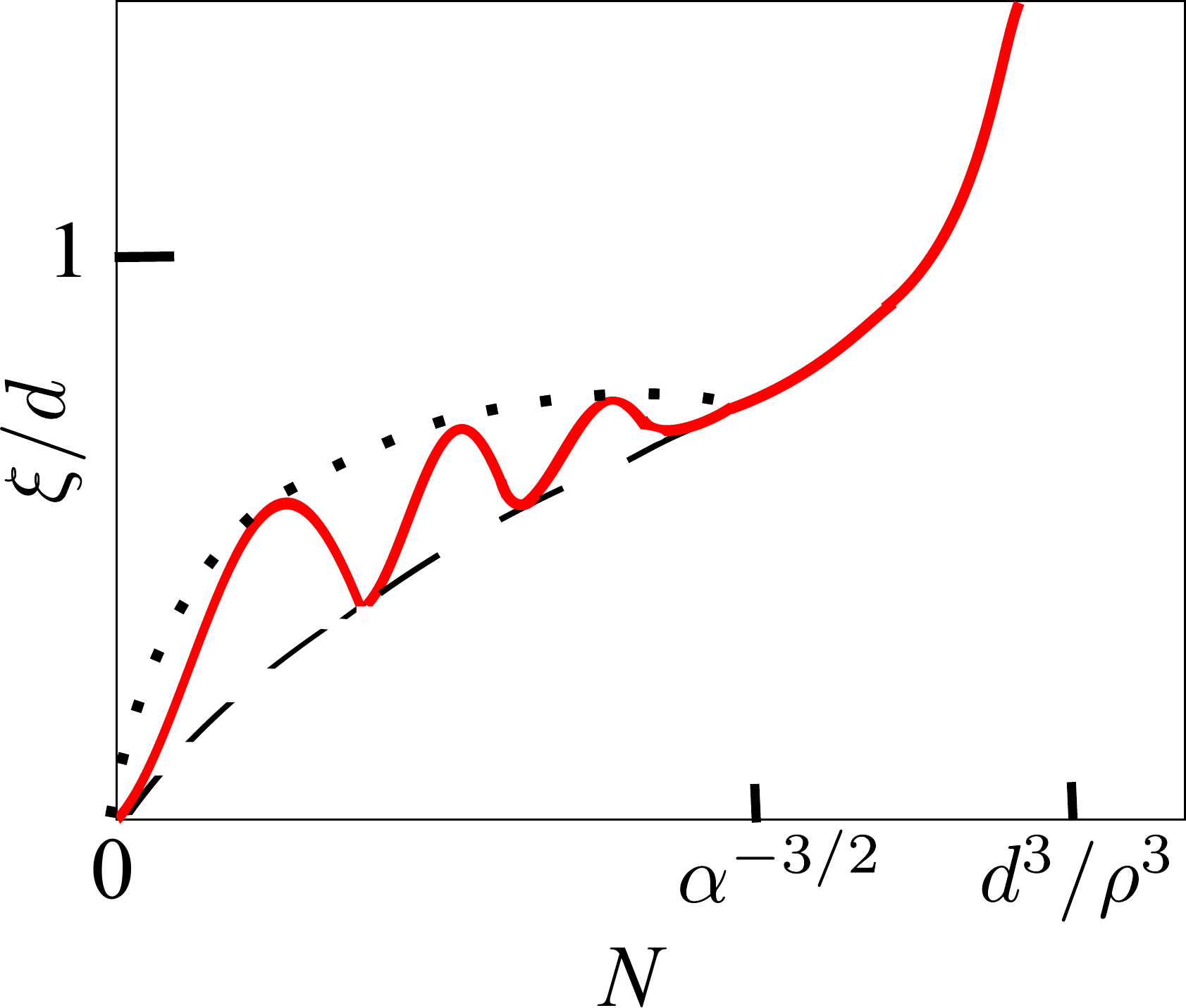}
\caption{(Color online) Schematic plot of the localization length $\xi$ (in units of the NC diameter $d$) as a function of the average number of donors $N$ in a NC at $\alpha d^2/\rho^2\gg 1$. The thick solid line (red) is the localization length. The dotted line represents the maximum value of $\xi$ given by Eq. \eqref{eq:max} which corresponds to the Fermi level position in the middle of the degenerate shell (the line of maxima). The dashed line goes through the minima of $\xi$ which are somewhat lower than values given by Eq. \eqref{eq:4} and happen when the Fermi level is near the middle of the gap $\Delta$. When $N=\alpha^{-3/2}$, the energy shift due to NC diameter variations $\gamma_1$ reaches $\Delta$, the oscillations of $\xi$ stop and $\xi$ obeys Eq. \eqref{eq:4}. At $N$ close to $d^3/\rho^3$, the film approaches the IMT and its localization length diverges.}\label{fig:oscil}
\end{figure}

So once the matrix element $t$ is known, we can get the localization length. For different types of contacts between NCs, the value of $t$ is different. The largest $\xi$ is obtained in the case when NCs touch by facets of finite radius $\rho$. The corresponding tunneling matrix is derived in Sec. \ref{sec:facet} as
\begin{equation}
t\simeq\frac{9\hbar^2 n\rho^3}{m^*d^2}.\label{eq:ft}
\end{equation}
The energy difference $\delta E$ oscillates with the density of states, which is followed by the oscillation of the localization length (see Fig. \ref{fig:oscil}). At small $N$ and when the Fermi level is inside a degenerate shell where the density of states is large, one arrives at the ES law \eqref{eq:ES} and gets $\delta E=\gamma_1\ll\Delta$. The localization length reaches a maximum at such $N$
\begin{equation}
\xi\simeq\frac{d}{\ln(\alpha d^2/n^{1/3}\rho^3)}.\label{eq:max}
\end{equation}
When the Fermi level resides in the middle of the gap $\Delta$ between shells where the Mott variable range hopping takes over, the energy difference $\delta E\simeq \Delta$. Therefore, the localization length reaches its periodic minima which are given by Eq. \eqref{eq:4}. The local period is $\sim N^{1/3}$ and slowly changes with $N$.
Eqs. \eqref{eq:max} and \eqref{eq:4} together give the envelope of the oscillating localization length as shown in Fig. \ref{fig:oscil} by the dotted line and the dashed line, respectively. We denote this phase as ``oscillating insulator" (OI). Periodic oscillations of the hopping conductivity with $N$ were observed in CdSe~\cite{Liu_2010}.

According to  Eq. \eqref{eq:y} $\gamma_1$ grows with $N$ and reaches $\Delta$
at $N=\alpha^{-3/2}$. At larger $N$ the energy difference $\delta E$ saturates at $\Delta$ because of the spectrum periodicity. The corresponding system of electron energy levels with a smooth density of states is shown in Fig. \ref{fig:dis}. Thus, oscillations of $\xi(N)$ stop at  $N=\alpha^{-3/2}$. We arrive at the usual insulator (I) where $\xi$ obeys Eq. \eqref{eq:4} which follows from $\delta E=\Delta$ and Eqs. \eqref{eq:x}, \eqref{eq:5} and \eqref{eq:ft}. This gives Eq. \eqref{eq:result} for the critical concentration $n_c$. This sequence of changes of $\xi(N)$ is shown schematically in Fig. \ref{fig:oscil}. Apparently it requires that
\begin{equation}
\alpha^{-3/2}\ll d^3/\rho^3. \label{eq:xx}
\end{equation}
In this section, we focus on the case of relatively small $\rho$ when the inequality \eqref{eq:xx} holds. In this case, every maximum of $\xi$ is finite because the argument of the logarithmic function in Eq. \eqref{eq:max} $\alpha d^2/n^{1/3}\rho^3=(\alpha d^2/\rho^2)^{3/2}/(N\alpha^{3/2})^{1/3}\gg 1$ at $N<\alpha^{-3/2}$. The case opposite to the inequality \eqref{eq:xx} corresponding to a smaller $\alpha$ and a larger $\rho$ is studied in the next section.

One should note that our result for $\xi$ is obtained away from the critical vicinity of $n_c$. So our estimate of $n_c$ obtained from the condition $\xi=d$ needs a correction. Indeed we estimated the probability of the electron hopping between two distant NCs via the elastic co-tunneling along a single typical chain of $M$ NCs. Near the IMT one should add probability amplitudes of many such chains. Then the sum of all amplitudes gives a total probability $\propto(tK/\Delta)^{M}$. Here $K$ is the connective constant of the NC lattice.
According to Anderson \cite{Anderson1972}, the IMT happens when $tK/\Delta =1$.
Using Eqs. \eqref{eq:x} and \eqref{eq:ft}, for the simple cubic lattice (where according to Ref. \onlinecite{Clisby} $K=4.7$) we arrive at the estimate $n_c \approx 0.5 \rho^{-3}$ while for the face-centered cubic lattice (where $K=10$ as given in Ref. \onlinecite{McKenzie}) we get $n_c\approx 0.2\rho^{-3}$. This result found from the insulating side of the IMT is reasonably close to Eq. \eqref{eq:result} obtained from the metallic side.

\begin{figure}[h]
\includegraphics[width=.45\textwidth]{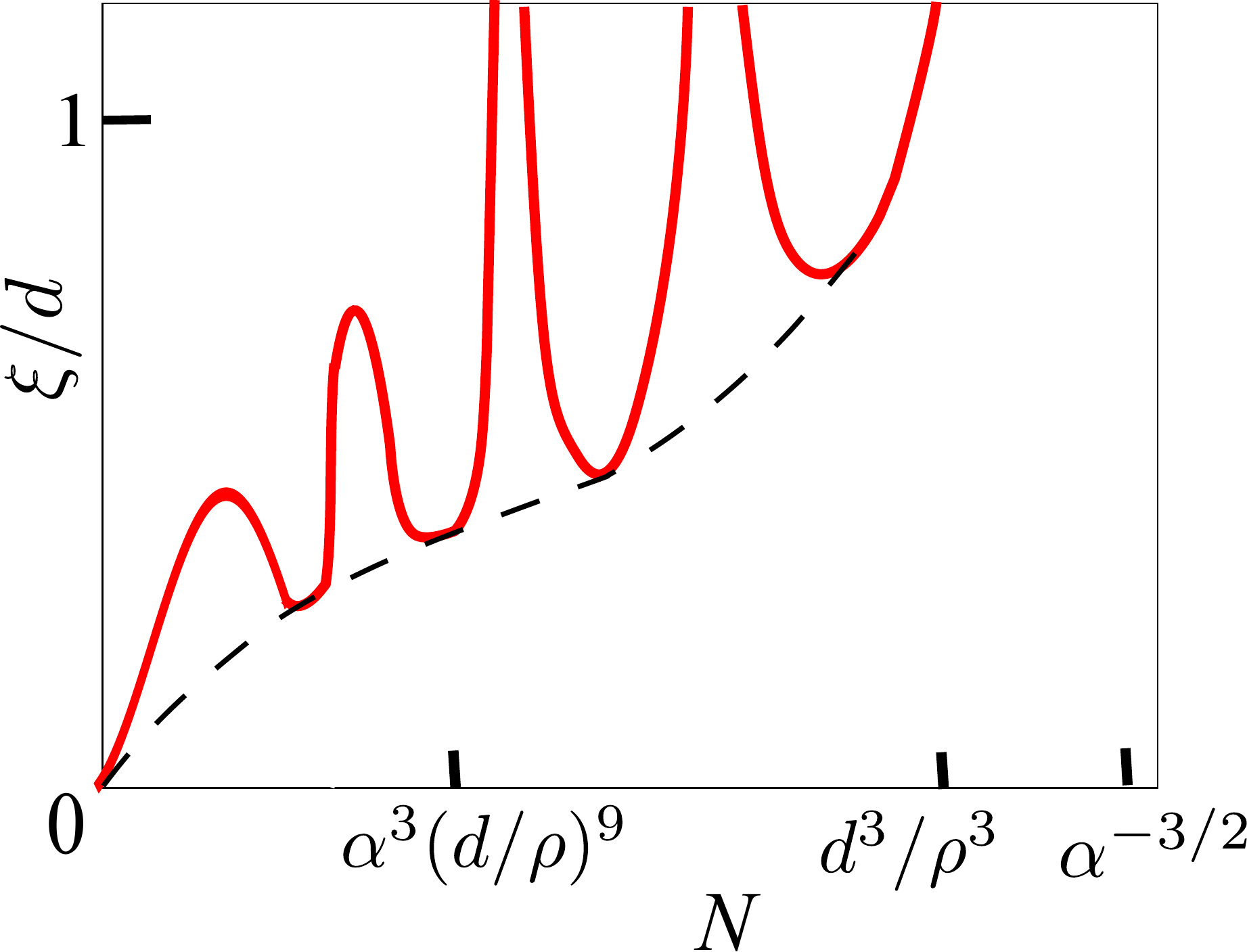}\\
\caption{(Color online) The localization length $\xi$ (in units of the NC diameter $d$ as a function of the average donor number $N$ in a NC for $\alpha d^2/\rho^2\ll 1$. The thick solid line (red) is the localization length. The dashed line represents the minima of $\xi$ somewhat lower than values given by Eq. \eqref{eq:4} occurring when the Fermi level is in the middle of the gap $\Delta$. The film first becomes a ``blinking metal" (BM) at $N=\alpha^3d^9/\rho^9$, where $\xi$ starts to periodically diverge and return to finite values. At $N=d^3/\rho^3$, the film enters the usual metal (M) domain.}\label{fig:osci2}
\end{figure}

So far, we have studied the case where the inequality \eqref{eq:xx} holds. Now we turn to the opposite situation $\alpha d^2/\rho^2\ll1$ of relatively large $\rho$ and small $\alpha$. In this case Eq. \eqref{eq:max} indicates that the localization length periodically diverges at $\alpha^3d^9/\rho^9<N< d^3/\rho^3$. This means that electrons whose energy levels are in the middle of the shell are delocalized while those who are located in the tails of the density of states are still localized and have a localization length described by Eq. \eqref{eq:4}. We call this phase ``blinking metal" (BM) since its metallicity occurs only at certain positions of the Fermi level (a good example of such metal is the quantum hall effect). However, at $N=d^3/\rho^3$, the system enters the usual metal (M) phase where electrons are delocalized regardless of the Fermi level position. The corresponding behavior of the localization length at $N\le d^3/\rho^3$ is shown in Fig. \ref{fig:osci2}. In this case, $\gamma_1\ll \Delta$ at the IMT point since $d^3/\rho^3\ll \alpha^{-3/2}$.

\section{Role of NC charging due to donor number fluctuations}
\label{sec:charge}
\begin{figure}[h]
\includegraphics[width=.45\textwidth]{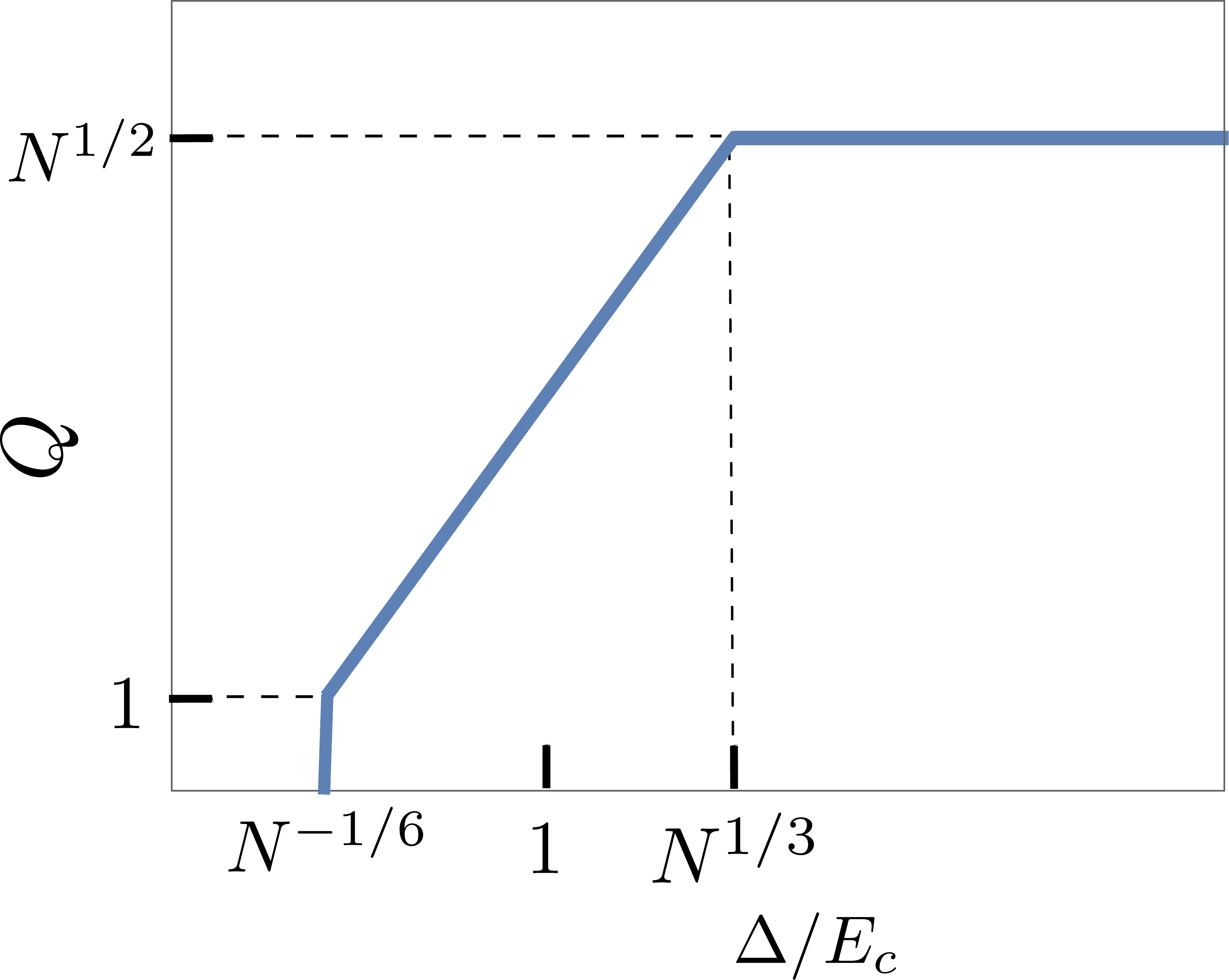}\\
\caption{Schematic log-log plot of the typical number of charges $Q$ in a NC as a function of the ratio $\Delta/E_c$.}\label{fig:charge}
\end{figure}

Let us now discuss another type of disorder present in the film, i.e., the fluctuations of the donor number $\delta N$ around the average number $N$ from NC to NC. At large $N$, $\delta N$ is Gaussian-distributed, i.e., $\delta N\sim\sqrt{N}$. If each NC is neutral, $\delta N$ would lead to substantial fluctuations $\delta E_F=E_F/\sqrt{N}\sim N^{1/6}\Delta$ of the Fermi energy $E_F$ from one NC to another. To establish the unique chemical potential of electrons (the Fermi level), electrons move from NCs with larger-than-average $n$ to ones with smaller-than-average $n$ and NCs get charged creating the Coulomb potential in space. Below we argue that the typical number of charges $Q$ in NCs depends on the ratio $\Delta/E_c$ as shown in Fig. \ref{fig:charge}.

When $E_c$ is very small, the final chemical potential is established when NCs have almost the same number of electrons. Accordingly, most NCs obtain a net charge $Qe$ where $Q\sim \sqrt{N}$. However, at larger $E_c$ when $\Delta/E_c\ll N^{1/3}$, the price of charging gets so large that the number of transferred electrons $Q\sim N^{1/6}(\Delta/E_c)$ is much smaller than $\sqrt{N}$ (see Fig. \ref{fig:charge}). One arrives to this result by equating the initial fluctuation of the Fermi energy $\delta E_F$ to the growth of the Coulomb potential of a NC $QE_c$. At $\Delta/E_c=N^{-1/6}$, charging becomes so costly that the charge number $Q=1$. Beyond this point, all NCs are neutral (see Fig. \ref{fig:charge}).

One can understand the importance of the parameter $\Delta/N^{1/3}E_c$ by calculating the electronic screening radius of the film. Since the screening radius $r_0$ can be estimated as $\sqrt{\varepsilon_f/4\pi e^2 g(E)}$ where $g(E)\simeq N^{1/3}/\Delta d^3$ is the average density of states, one gets $r_0/d\simeq\sqrt{\Delta/N^{1/3}E_c}\gg1$ at $\Delta/N^{1/3}E_c\gg1$.
We see that in agreement with Fig. \ref{fig:charge}, when $\Delta/N^{1/3}E_c\gg 1$  and $r_0 \gg d$
electrons do not screen donor charges, while in the opposite case $\Delta/N^{1/3}E_c \ll 1$ the electron screening becomes important.

Due to charging of NCs, each NC finds itself in the environment of charged neighbors and gets a random potential energy shift up or down. Apparently the energy shift created by a single NC at the distance $r$ where $d \ll r \ll r_0$ is $QE_cd/r$
and the typical shift created collectively by all NCs in the sphere of radius $r_0$ is
\begin{equation}
\gamma_2 = \frac{QE_cd}{r_0}\left(\frac{r_0}{d}\right)^{3/2} =N^{5/12} E_c \left(\frac{\Delta}{E_c}\right)^{1/4}.\label{eq:z}
\end{equation}
Note that $\gamma_2$ does not depend on $d$. This is not surprising because one can arrive to the same result for potential energy fluctuations thinking about our film as a bulk heavily doped semiconductor with concentration $n \simeq N/d^{3} $ of randomly positioned donors screened by degenerate electron gas~\cite{Shklovskii}.

\begin{figure}[h]
\includegraphics[width=.47\textwidth]{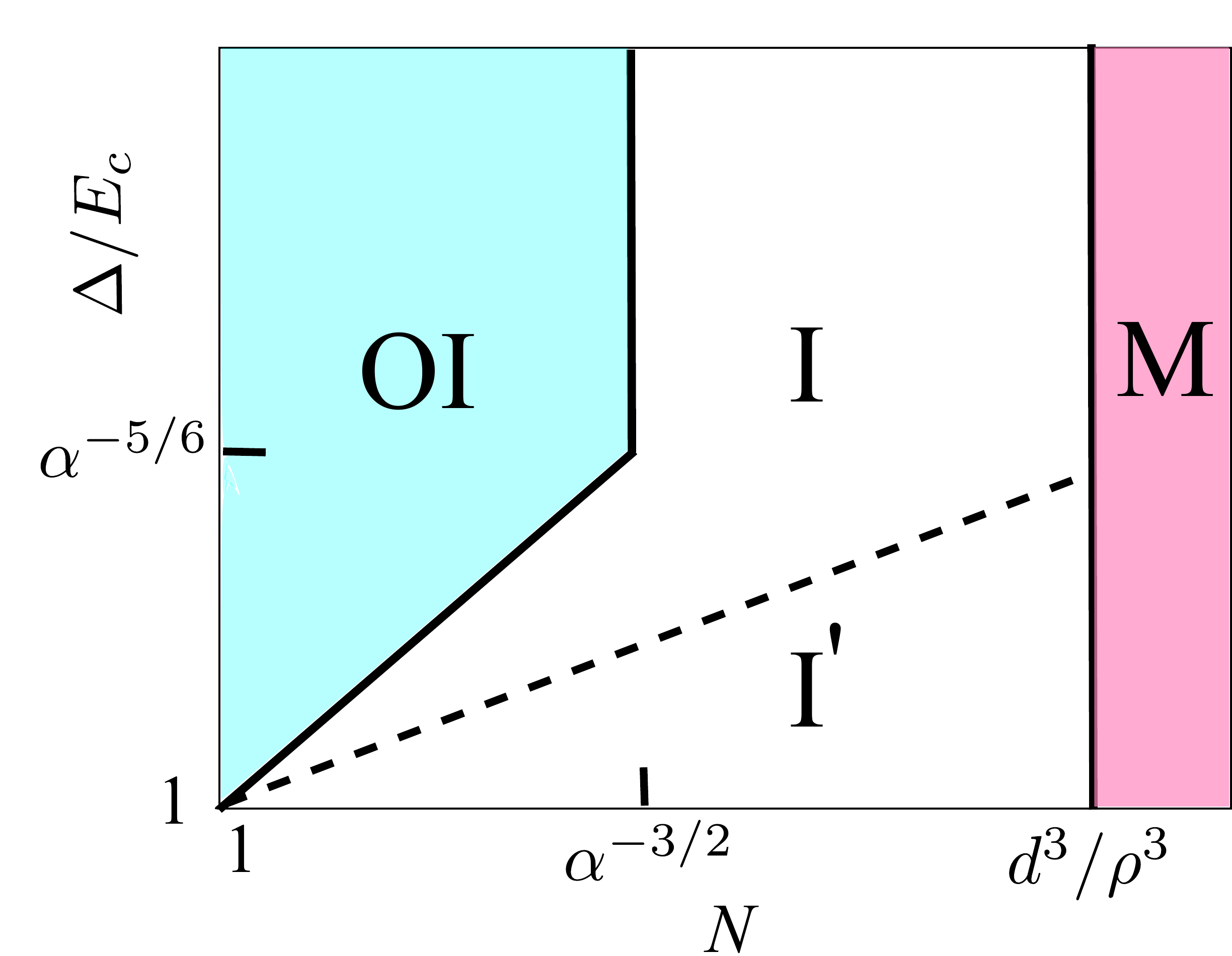}
\caption{(Color online) A log-log map of different domains in the ($N$, $\Delta/E_c$) plane at $\alpha d^2/\rho^2\gg 1$. The light grey domain (blue online) corresponds to the regime where $\delta E<\Delta$ and the localization length oscillates with $N$. We call it an oscillating insulator (OI). Away from the left blue domain, the energy difference $\delta E$ saturates at $\Delta$ and the localization length does not oscillate and obeys Eq. \eqref{eq:4}. This is the usual insulator (I). The darker grey domain (pink online) corresponds to the metallic phase (M). The solid border lines (from left to right) correspond to equations $\Delta/E_c=N^{5/9}$, $N=\alpha^{-3/2}$ and $N=d^3/\rho^3$, respectively. The dashed line corresponds to $\Delta/E_c=N^{1/3}$. In the domain below the dashed line (I'), $(n,l)$-shells get split and mixed with each other (see Sec. \ref{sec:nond}). In this domain, the localization length is given by Eq. \eqref{eq:deg}.}\label{fig:lowN}
\end{figure}

Let us find what happens when the charging effect outweighs the diameter variation. We start from the case $\alpha d^2/\rho^2\gg 1$ and use that for Fig. \ref{fig:lowN} which in the $(N, \Delta/E_c)$ plane shows phases with different behaviors of the localization length. The upper part of Fig. \ref{fig:lowN} summarizes results obtained for diameter variations in Sec. \ref{sec:NCd}.

We see how with growing $N$ the film goes through an oscilating insulator (OI), an insulator (I) and a metal (M). Coulomb effects become important when $\gamma_2>\gamma_1$ or according to Eqs. \eqref{eq:y} and \eqref{eq:z} $\Delta/E_c<1/N^{1/3}\alpha^{4/3}$. At the upper OI-I border $N = \alpha^{-3/2}$ where $\gamma_1=\Delta$, this happens at $\Delta/E_c = \alpha^{-5/6}$. Let us explore now what happens at $\Delta/E_c < \alpha^{-5/6}$ where the energy difference $\delta E=\gamma_2$. When $N$ is small, $\gamma_2$ is small too so that the density of states has periodic peaks and the localization length oscillates. The system is again an oscillating insulator (OI) (see the narrower part of the blue domain in Fig. \ref{fig:lowN}). At larger $N$ when $\Delta/E_c< N^{5/9}$, the energy shift $\gamma_2$ exceeds $\Delta$ so that away from the left blue domain shown in Fig. \ref{fig:lowN} we arrive at the spacial distribution of levels shown in Fig. \ref{fig:dis}. $\delta E$ then saturates at $\Delta$ and one again obtains the result \eqref{eq:4} for the localization length $\xi$. The system becomes a usual insulator (I). Thus in the case of relatively small $\rho$ and large $\alpha$ when $\alpha d^2/\rho^2\gg 1$, the localization length first stops oscillating and then diverges (the system enters first the domain I and then the domain M as shown in Fig. \ref{fig:lowN}).

\begin{figure}[h]
\includegraphics[width=.47\textwidth]{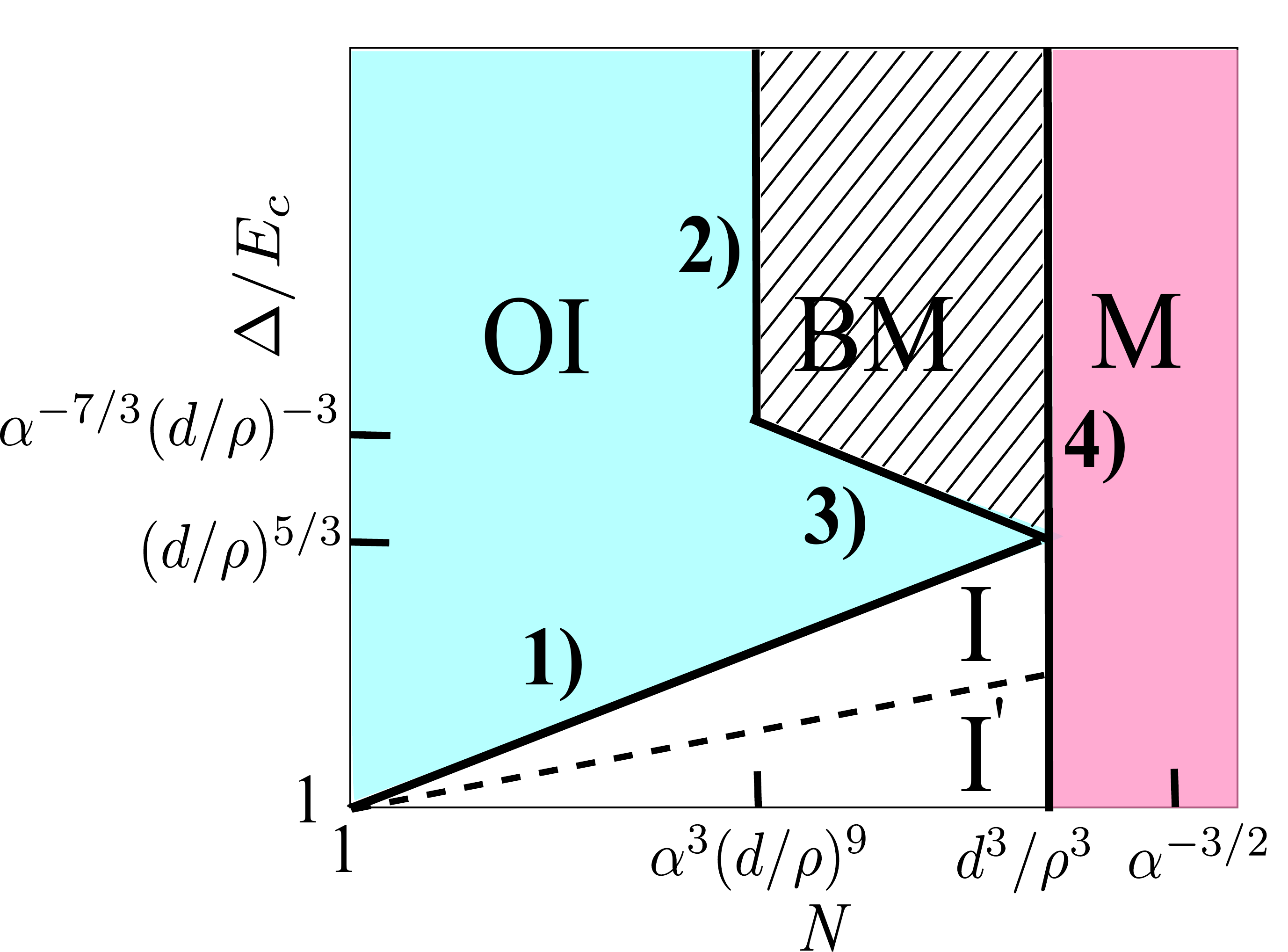}\\
\caption{(Color online) A log-log map of different domains in the ($N$, $\Delta/E_c$) plane at $\alpha d^2/\rho^2\ll1$. The light grey domain (blue online) corresponds to the regime where $\delta E<\Delta$ and the localization length oscillates with $N$ (OI). The shaded domain belongs to the blinking metal (BM). In the white domain of the usual insulator (I) the localization length does not oscillate and obeys Eq. \eqref{eq:4}. The darker grey domain (pink online) corresponds to the metallic phase (M). The solid border lines 1)--4) correspond to conditions 1) $\gamma_2=\Delta$, 2) $\gamma_1=t$, 3) $\gamma_2=t$, and 4) $\Delta=t$. The dashed line corresponds to $\Delta/E_c=N^{1/3}$. In the domain I' below the dashed line NC spectra become random (see Sec. \ref{sec:nond}). In this domain, the localization length is given by Eq. \eqref{eq:deg}.}\label{fig:final2}
\end{figure}

In the opposite case of relatively large $\rho$ and very small $\alpha$ where $\alpha d^2/\rho^2\ll 1$ we analyze the role of Coulomb effects in Fig. \ref{fig:final2}. The upper part of this phase diagram $\Delta/E_c>\alpha^{-7/3}(d/\rho)^{-3}$ is again dominated by diameter variations. As shown in Sec. \ref{sec:NCd} in this case with growing $N$ the film goes through an oscillating insulator (OI), a blinking metal (BM) and a metal (M). When we include Coulomb effects the vertical OI-BM border marked as the line 2) in Fig. \ref{fig:final2} at which $\gamma_1= t$ cannot continue below the point $\Delta/E_c=\alpha^{-7/3}(d/\rho)^{-3}$ where $\gamma_1=\gamma_2$. Now the OI-BM border in Fig. \ref{fig:final2} turns and goes along the line 3) at which $\gamma_2 =t$ or $\Delta/E_c=(d/\rho)^4/N^{7/9}$. The line 3) ends at $\Delta/E_c=(d/\rho)^{5/3}$ when $\gamma_2= \Delta$ at the crossing with the generic metal border given by the line 4) where $\Delta=t$ and $N=d^3/\rho^3$ and with the OI-I border given by the line 1) where $\gamma_2 = \Delta$ and $\Delta/E_c=N^{5/9}$. This is a remarkable quadruple point where $\gamma_1=\gamma_2=\Delta =t$ and all four phases OI, BM, I and M meet.

\section{Experimental implications for $\mathbf{CdSe,\, InAs}$ and $\mathbf{ZnO}$ NC}
\label{sec:discussion}

In previous section, we have studied theoretically possible situations for NC films as shown by phase diagrams in Figs. \ref{fig:lowN} and \ref{fig:final2}. Now we focus on several commonly used semiconductor NCs trying to put them on these diagrams. We choose the same geometrical parameters $\alpha=0.15$, $d=5$ nm, $\rho=\rho_a=1.2$ nm for all of them. Then, we get $\alpha^{-3/2}\lesssim d^3/\rho^3$ and use the phase diagram Fig. \ref{fig:lowN}. The upper part of the diagram where the NC diameter variation is the major source of disorder is separated from the lower one where Coulomb disorder dominates by $\Delta/E_c=\alpha^{-5/6}\simeq 5$.  For CdSe NCs since $\Delta/E_c = 5$, the Coulomb effects are marginal so that we can think about the NC diameter variation only. When  $N$ increases, the film moves from OI to M as depicted by the upper part of the phase diagram in Fig. \ref{fig:lowN} with the intermediate region I being narrow and neglected since $\alpha^{-3/2}$ and $d^3/\rho^3$ are quite close. For InAs, since $\Delta/E_c=27\gg 5$, the Coulomb effects are completely negligible and again the system experiences the OI-I-M phase changes. For ZnO, however, the ratio $\Delta/E_c=3 < 5$ and the random Coulomb potential is the leading disorder in the film. One should then use the lower part of Fig. \ref{fig:lowN} for the phase change process with increasing $N$. In this case we have the same sequence of phases OI-I-M but with now the phase I appreciably expanded by the Coulomb random potential. Si NC films are similar to that of ZnO as its $\Delta/E_c=2$ is also very small.

One should note that there is no BM phase for the chosen parameters. To get this phase, one has to tune $\alpha$ down by making the NCs more monodisperse.  For $d=5$ nm and $\rho=\rho_a=1.2$ nm, one needs $\alpha<0.06$ to open the BM phase. This is probably the state-of-the-art monodispersity. To go even further, one may wonder whether the BM phase can be expanded all the way till $N=1$. The inequality $\alpha d^3/\rho^3\le1$ guarantees that the line 2) of Fig. \ref{fig:final2} reaches the $N=1$ line and simultaneously the condition $\Delta/E_c \geq (d/\rho)^{4}$ is required for the film to be above the point where line 3) crosses $N=1$. If both inequalities are satisfied one can expect a desirable \cite{shabaev_dark_2013,Kagan,Yang_2015} band-like transport behavior of electrons when they populate only the 1S-level. However, for NCs with $d=5$ nm and $\rho=\rho_a\simeq 1.2$ nm, the necessary $\alpha=\rho^3/d^3\simeq 0.01$ is unrealistically small while necessary  $\Delta/E_c \geq (d/\rho)^{4} \sim 200$ is too large. Even increasing $\rho$ to 2$\rho_a$  brings us only to criteria $\alpha \leq 0.1$ and  $\Delta/E_c \geq 16$. Of course, our estimates are good only for $\rho \ll d/2$ so these numbers should not be taken too seriously. For ZnO (or Si), since $\Delta/E_c<(d/\rho)^{5/3}$ even at $\rho=2$ nm, the system can never see a BM phase due to the large Coulomb disorder as shown by Fig. \ref{fig:final2}.

There is an important case where additional Coulomb fluctuations may be ignored. We are talking about NC films gated by an ionic liquid or an electrolyte~\cite{Liu_2010,Uttrecht}. Anions which enter spaces between NCs and attract electrons in this case play the role of chemical donors we studied above. However, contrary to immobile dopants inside a NC anions remain mobile in the process of adjustment of the gate voltage and tend to screen electron charges~\cite{Boris}. Thus, in this case disorder effects due to fluctuations of NC diameters discussed in Sec. \ref{sec:NCd} should dominate. The ZnO (or Si) NC films then become similar to CdSe or InAs. At $\alpha d^2/\rho^2\gg 1$ the OI domain expands while the I domain shrinks and at $\alpha d^2/\rho^2\ll 1$, the OI and I regions are consumed by the BM phase.

\section{Tunneling matrix element for nanocrystals touching by facets}
\label{sec:facet}

\begin{figure}[h]
$\begin{array}{c}
\includegraphics[width=0.5\textwidth]{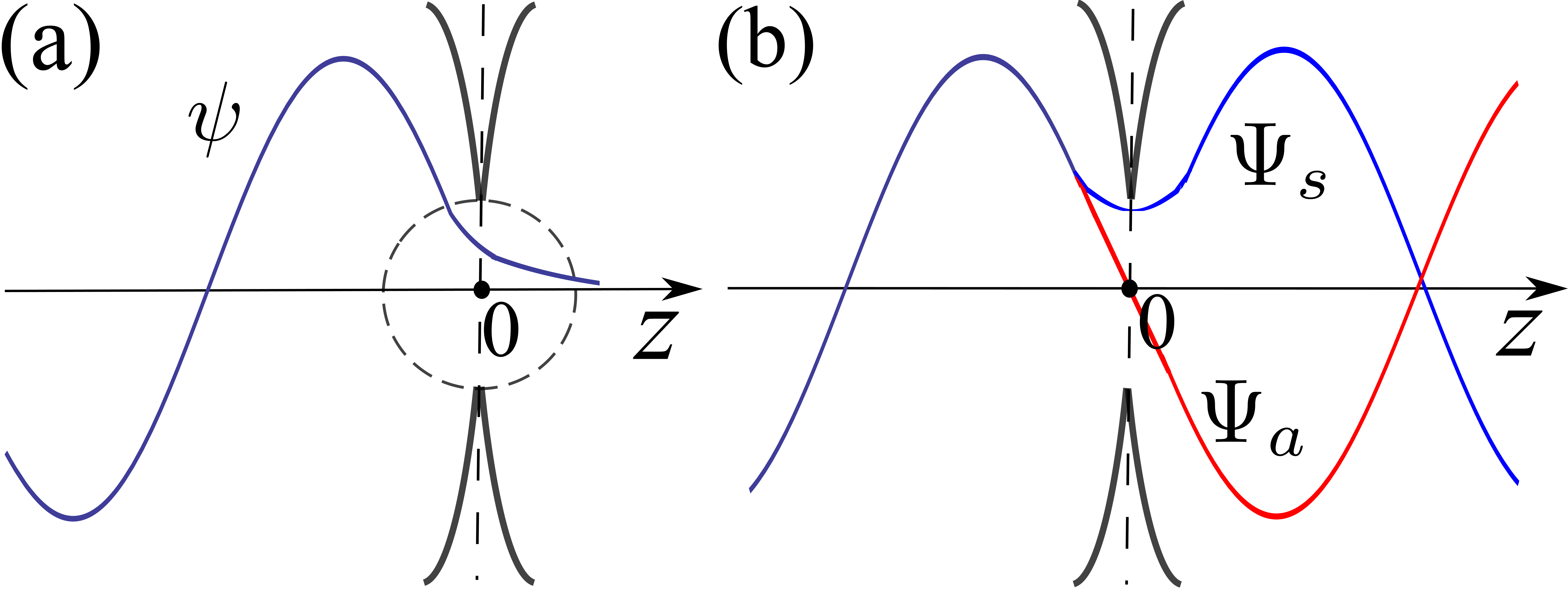}
\end{array}$
\caption{(Color online) Electron wave functions near the contact facet. The vertical dashed line indicates the facet boundary plane. (a) Electron wave function $\psi$ of the left NC modified by the right one. It acquires a tail penetrating into the right NC mainly in the region of the dashed sphere. (b) Overall wave function as symmetric (blue) or asymmetric (red) combinations of each NC wave function given by Eq. \eqref{eq:sym}. The energy difference between these two states is twice the tunneling matrix element $t$ given by Eq. \eqref{eq:9}.}\label{fig:2}
\end{figure}

Beyond the surface of a single NC in the surrounding medium, the wave function of an electron at the Fermi level decays with the distance $s$ from the surface as $\propto e^{-s/b}$ where $b=\hbar/\sqrt{2mU_0}$. Here $m$ is the electron mass in the medium and $U_0$ is the workfunction of NCs. For $U_0\simeq 4$ eV and $m=m_e$, where $m_e$ is the electron mass in vacuum, one gets $b\simeq1\AA$, which is smaller than the lattice constant. So, approximately, the electron wave function is zero on the surface of NCs. When two NCs touch by contact facets, the electron wave function of the left NC is strongly modified inside the dashed sphere of radius $\rho$ containing the facets. Namely, due to the right NC, the wave function acquires a tail leaking into the right NC (see Fig. \ref{fig:2}a). 
The wave function inside the right NC is deformed in the same way.
The overall wave function is split into two
\begin{equation}
\label{eq:sym}
\Psi_{s,a}(\boldsymbol{r})=\frac{1}{\sqrt{2}}\left[\psi(\boldsymbol{r}-\boldsymbol{r_L})\pm\psi(\boldsymbol{r}-\boldsymbol{r_R})\right],
\end{equation}
which are symmetric and asymmetric combinations of the modified wave function $\psi$ inside each NC (see Fig. \ref{fig:2}b). The origin is set at the center of the contact and the polar axis is pointed towards the center of the right NC. The coordinates of the centers of left and right NCs are $\boldsymbol{r_L}$ and $\boldsymbol{r_R}$, respectively. $\psi(\boldsymbol{r}-\boldsymbol{r_L})$ refers to the wave function in the left NC and $\psi(\boldsymbol{r}-\boldsymbol{r_R})$ is that of the right one. Below, we just use the left wave function for discussion and simply denote it as $\psi$. The tunneling matrix element $t$ between two NCs can be estimated by calculating the energy splitting between the symmetric and asymmetric wave functions $\Psi_{s,a}$ of Eq. \eqref{eq:sym}. As in the problem of calculating the electron terms of the molecular ion H$_2^+$ in $\S$ 81 of Ref. \onlinecite{Landau}, the energy splitting can be calculated as
\begin{equation}
t=\int \frac{\hbar^2}{m^*}\psi\frac{d\psi}{dz}dx\,dy\label{eq:9}
\end{equation}
where the integral is taken over the contact boundary plane $z=0$ (see the vertical dashed line in Fig. \ref{fig:2}b). In the case we are discussing now, the contact is made of the touching facets. In this contact plane, $\psi$ vanishes at $(x,y)$ outside the facets in the surrounding medium.

Let us deal with $E_F$ belonging to a degenerate $(n,l)$-shell. Then, the unperturbed wave function of the left NC is $\psi_0(\boldsymbol{r}-\boldsymbol{r_L})\simeq j_l(k_nr')Y_l^m(\theta',\phi')/\sqrt{d/k_n^2}$ where $j_l$ is the spherical Bessel function, $Y_l^m$ are spherical harmonics, $k_nd/2$ is the $n$th zero point of the Bessel function and $k_n\approx 2\pi n/d\sim k_F$, $(r',\,\theta',\,\phi')$ are the coordinates of $\boldsymbol{r'}=\boldsymbol{r}-\boldsymbol{r_L}$ in the spherical coordinate system. $Y_l^m(\theta',\phi')\rightarrow 0$ at $\theta'\rightarrow 0$ for all $m\ne 0$, and for $m=0$ $Y_l^0(0,\phi')=\sqrt{(2l+1)/4\pi}>0$. Thus among the $2l+1$ degenerate levels of the $(n,l)$-shell only one state ($m=0$) oriented along the $z$ axis contributes to the tunneling between neighboring NCs (marked red in Fig. \ref{fig:dis}). So we just need to calculate the tunneling matrix element $t$ of the $m=0$ state. When the number $N$ of electrons inside each NC is large, for the $(n,l)$-shell at the Fermi level, we have $n\sim l\sim N^{1/3}\simeq k_Fd$ since the radial and angular kinetic energies should be of the same order. So for the $m=0$ state, the wave function is highly concentrated near the $z$ axis spreading mainly within the polar angle $\simeq1/\sqrt{k_Fd}$. More accurately, since each $(n,l)$-shell has $2l+1$ degeneracy, we get
\begin{equation}
N\approx2n(l+1)^2\sim2l^3,
\end{equation}
where the factor 2 comes from the spin degeneracy. As $N=4\pi n(d/2)^3/3=k_F^3d^3/18\pi$, we get $
n\sim l\sim\left(k_F^3d^3/36\pi\right)^{1/3}$.
The radial distribution is described by $j_l(k_nr')\simeq\sin[k_n(r'-d/2)]/k_nr'$ at large $r'$. Therefore, we get approximately the normalized unperturbed wave function
\begin{equation}
\psi_0\approx\frac{2\sin\left[k_n(r'-d/2)\right]}{\sqrt{d}r'}Y_l^0(\theta',\phi')
\end{equation}
at large distance from the left NC center. So near the facet, the original unperturbed wave function $\psi_0$ can locally be regarded as an incident plane wave superposed by its completely reflected wave from the surface, i.e., $\psi_0\approx2\sqrt{2l/\pi}\sin[k_n(z'-d/2)]/d^{3/2}$, where $z'\approx r'$ is the $z$-component of $\boldsymbol{r'}$.

As a result, the problem of an electron tunneling through a facet is analogous to the one of a plane wave with the wavenumber $k_F$ diffracting on a circular aperture with radius $\rho$ in $z=0$ plane screen. In the regime where $k_F \rho \ll 1$,  Bethe \cite{Bethe} solved this diffraction problem for microwaves, while Levine and Schwinger \cite{Schwinger} and Bouwkamp \cite{Bouwkamp} solved it for a scalar plane wave. Here we use the simple solution in the first-order approximation in $k_F \rho\ll1$ given by Rayleigh \cite{Rayleigh}.

One can write the Schrodinger equation for the function $\psi$ as

\begin{equation}
  \label{eq:Schrodinger}
  \nabla^2 \psi + k_F^2 \psi=0.
\end{equation}
\noindent Boundary conditions on the $z=0$ plane are $\psi=0$ on the screen and $d\psi/dz$  is continuous at the aperture. We write the solution  $\psi$ as the sum of $\psi_0$ and $\delta\psi$, where $\delta\psi$ is the correction due to the aperture opening and the unperturbed wave function $\psi_0$ of the left NC is zero on the right side of the boundary plane ($z>0$). We denote $\delta\psi_L,\delta\psi_R$ as the left ($z<0$) and right ($z>0$) part of the correction function $\delta\psi$ respectively. So in the $z=0$ boundary plane, $\delta\psi_L=\delta\psi_R$ and outside the aperture $\delta\psi_L=\delta\psi_R=0$. The continuity of the derivative $d\psi/dz$ leads to a jump of $d\left(\delta\psi\right)/dz$, i.e., $d\psi_0/dz+ d(\delta\psi_L)/dz=d(\delta\psi_R)/dz$. The symmetry between $\delta\psi_L$ and $\delta\psi_R$ gives $d(\delta\psi_L)/dz=-d(\delta\psi_R)/dz$ in the aperture (the proof can be found in Refs. \onlinecite{Bethe,Rayleigh}. For a possible interpretation of this result, see the footnote \footnote{Let us consider the standing-wave solution of Shrodinger equation Eq. \eqref{eq:Schrodinger} in a free space: $\psi=\exp(ik_Fz)-\exp(-ik_Fz)$. Since $\psi=0$ at $z=0$, this means that $\psi$ is also the solution of the Shrodinger equation if the screen is located at $z=0$. This solution can be viewed as two plane waves $\exp(ik_Fz)$ and $-\exp(-ik_Fz)$, which fall from opposite directions on the screen and are reflected back completely. The aperture gives rise to corrections $\delta \psi_R$ and $\delta \psi_L$ to each wave. Since the screen does not affect the wave function $\psi$, these corrections to plane waves cancel each other, i.e. $\delta\psi_R(z)-\delta\psi_L(-z)=0$. By differentiating this relation we get that $d(\delta \psi_R)/dz=-d(\delta\psi_L)/dz$}) and therefore $d(\delta\psi_R)/dz=(d\psi_0/dz)/2\approx\sqrt{2l/\pi}k_n/d^{3/2}$. Now one can rewrite the integral for $t$ in terms of the correction to the wave function on the right side which is $\delta\psi_R$
\begin{equation}
  \label{eq:t_psir}
  t=\frac{\hbar^2}{m^*} \int  \delta\psi_R \frac{d\delta\psi_R}{dz}  dx dy,
\end{equation}
where $\delta\psi_R$ satisfies the Schrodinger equation (\ref{eq:Schrodinger}). At the aperture $\nabla^2(\delta \psi_R )\sim \delta\psi_R/\rho^2\gg k_F^2 \delta\psi_R$ because $k_F\rho \ll 1$. In the first approximation we can neglect the latter term and thus deal with the Laplace's equation \begin{equation}
\label{eq:Lap}
\nabla^2 (\delta\psi_R)=0
\end{equation}
with the boundary conditions $\delta\psi_R=0$ on the screen and $d(\delta\psi_R)/dz\approx\sqrt{2l/\pi}k_n/d^{3/2}$ at the aperture.

Mathematically, an identical problem was exactly solved in hydrodynamics (see  $\S$ 108 in Ref. \onlinecite{Lamb}). Indeed, the Laplace's equation $\nabla^2\varphi=0$ can be used to describe the motion of a rigid disk of radius $\rho$  moving with velocity $u$ along its axis (defined as the $z$ axis with the origin at the disk center) through unlimited incompressible liquid if $\varphi$ denotes the velocity potential. Boundary conditions for $\varphi$ are that $\nabla\varphi=u$ on the disk and $\varphi=0$ at $z=0$ outside the disk. The kinetic energy of the liquid in the $z>0$ space is

\begin{equation}
  \label{eq:kinetic}
  K=\frac{1}{2} g \int \left(\nabla\varphi\right)^2 dV,
\end{equation}
where $g$ is the density of the liquid. Using Green's theorem and the Laplace's equation for the right half space ($z>0$), we get
\begin{equation}
  \label{eq:kinetic_2}
  K=\frac{1}{2} g \int \varphi\frac{d \varphi}{d z} dxdy
\end{equation}
where the integral is taken over the whole $z=0$ plane. The potential $\varphi$ is zero outside the disk. Therefore, the integration is over the disk only, as in Eq. \eqref{eq:t_psir}.
Knowing the exact solution for $\varphi$ one can arrive at $K = (2/3) g \rho^3 u^2$ (see $\S$ 108 in Ref. \onlinecite{Lamb}). Thus
\begin{equation}
\label{eq:int}
\int \varphi\frac{d \varphi}{d z} dxdy=\frac{4}{3}\rho^3u^2.
\end{equation}
In our diffraction problem, $\delta\psi_R$ plays the role of $\varphi$ and $d(\delta\psi_R)/dz\approx \sqrt{2l/\pi}k_n/d^{3/2}$ plays the role of $u$. Therefore, using Eq. \eqref{eq:Fermi_wave}, we get
the tunneling matrix element $t$ in Eq. \eqref{eq:t_psir}
\begin{equation}
\label{eq:overlap}
t =\frac{9\hbar^2n\rho^3}{m^*d^2}= \frac{0.3\hbar^2 k_F^3 \rho^3}{m^*d^2}.
\end{equation}
At $k_Fd\sim 1$, one gets the tunneling matrix element for the 1s band
\begin{equation}\label{eq:gs}
t\simeq\frac{\hbar^2\rho^3}{m^*d^5}\,\,.
\end{equation}
In Ref. \onlinecite{Efros}, a solution-based oriented attachment method was used to prepare fused dimers of two semiconductor NCs. These dimers can be seen as two NCs touching by their facets. Eq. \eqref{eq:gs} for $t$ can then be used to calculate the splitting of the first exciton absorption line in the dimer spectrum. One should note that Eq. \eqref{eq:gs} is obtained here in the limit of infinitesimal tunneling distance $b$ (which is further explained in Sec. \ref{sec:bcon}). In the same limit, the method used in Ref. \onlinecite{Efros} leads to a smaller $t\simeq \hbar^2\rho^4/m^*d^6$. The reason for this difference is that on the facet plane our wave function has a larger magnitude than the one conjectured in Ref. \onlinecite{Efros}.

One can interpret the result for the tunneling matrix element Eq. \eqref{eq:overlap} as following. Originally the wave function $\psi_0$ is zero on the boundary plane and its derivative along the $z$ axis is $\simeq k_F^{3/2}/d$ on the contact facet. Now due to the existence of the facet, the electron wave function is modified as $\psi$ which leaks into the right NC and is nonzero on the facet, while the derivative is hardly changed by the small perturbations. Because the wave function substantially changes over a distance $\rho$, we can say that $\psi\approx (d\psi/dz )\rho$. As a result,
\begin{equation}
\begin{aligned}\label{eq:fac}
\psi\simeq&\frac{k_F^{3/2}\rho}{d},\\
\frac{d\psi}{dz}\simeq&\frac{k_F^{3/2}}{d}
\end{aligned}
\end{equation}
inside the contact facet in the $z=0$ plane for the $m=0$ state which is highly oriented along the $z$ axis. So we get the result \eqref{eq:overlap} for $t$. From this $t$ we arrive at Eq. \eqref{eq:4} for the localization length and Eq. \eqref{eq:result} for the critical concentration $n_c$.

\begin{figure}[h]
\includegraphics[width=.47\textwidth]{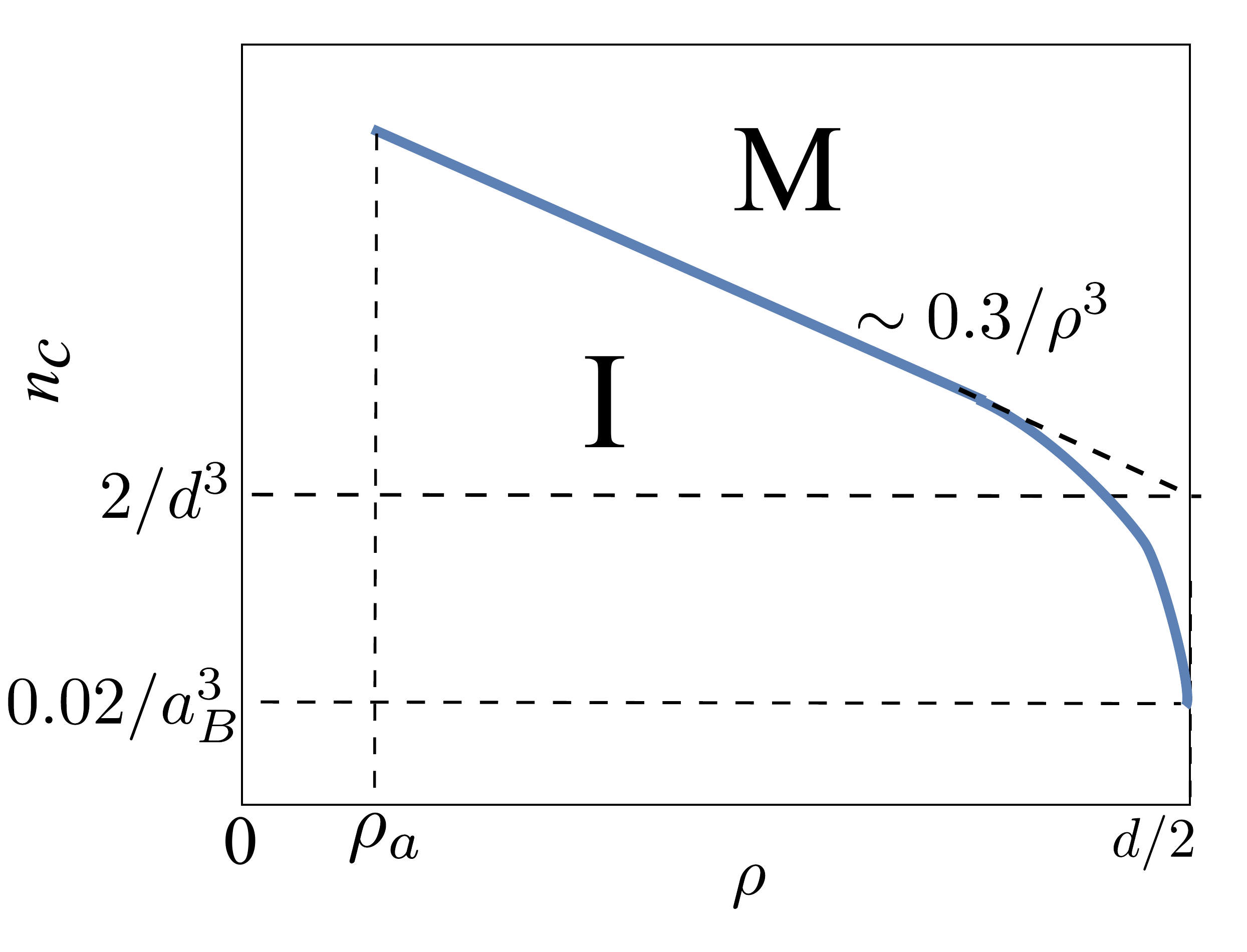}\\
\caption{Schematic logarithmic plot of the critical concentration $n_c$ as a function of the facet radius $\rho$ at $a_B\gg d$. Both axes use logarithmic scales. Near $\rho=d/2$, the critical concentration abruptly drops to its value for the bulk semiconductor.}\label{fig:5}
\end{figure}

A schematic plot of $n_c$ as a function of the facet radius $\rho$ is presented in Fig. \ref{fig:5}. The critical concentration scales as $\simeq0.3/\rho^3$ at all $\rho\ll d/2$. In the vicinity of $\rho=d/2$, electrons are no longer confined inside each NC and the film becomes a bulk semiconductor. In this case, $n_ca_B^3\simeq 0.02$, we return to the Mott criterion for the IMT and get a drastic drop of the critical concentration from $\simeq 2/d^3$ to $0.02/a_B^3$.

\section{Nanocrystals touching away from facets}
\label{sec:bcon}

\begin{figure}[h]
\includegraphics[width=0.45\textwidth]{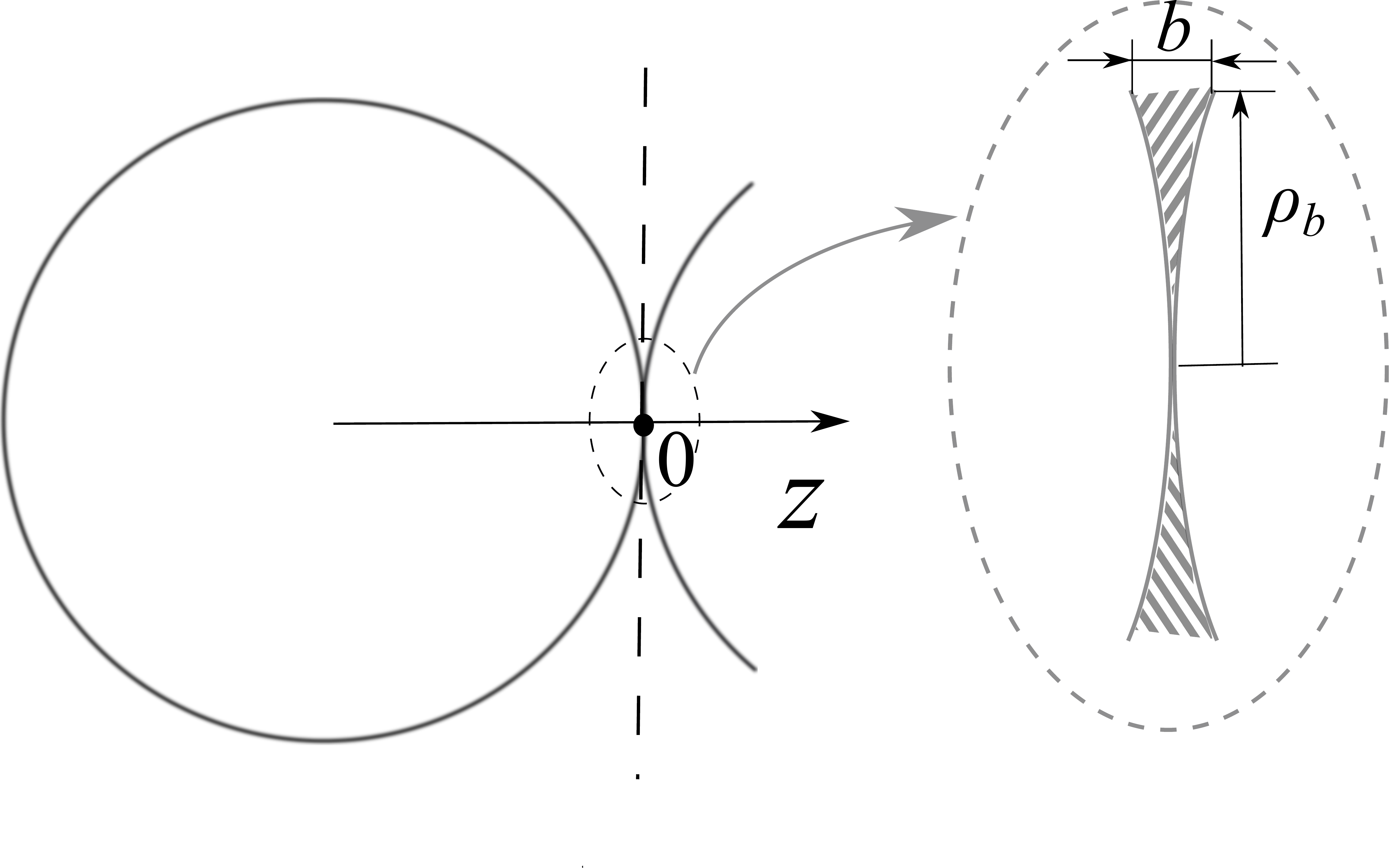}\\
\caption{Two NCs touching away from prominent facets. In this case, electrons tunnel through the $b$-contact shown in the inset.}\label{fig:3}
\end{figure}

When NCs touch each other away from prominent facets by an area of the atomic size $a\ll\rho$, the electrons tunnel mainly via the effective ``$b$-contact" of radius $\rho_b=\sqrt{db/2}\gg a$ (see Fig. \ref{fig:3}). For electrons tunneling between NCs outside this contact, the tunneling distance is larger than $b$ and the probability is negligible due to the exponentially decaying wave function.
Since electrons have to tunnel through the medium where they have mass $m$, when calculating the integral Eq. (\ref{eq:9}) over the contact boundary plane here, we need to replace the effective mass $m^*$ with $m$
\begin{equation}
t=\int \frac{\hbar^2}{m}\psi\frac{d\psi}{dz}dx\,dy.\label{eq:20}
\end{equation}
In this case we can use the LCAO approximation in the way done for the ground state in Ref. \onlinecite{shabaev_dark_2013}. We calculate $\psi=\psi_0$ as the wave function for a single spherical NC embedded in the infinite surrounding medium with the finite decay length $b$. For simplicity, in this section and below we do only scaling analysis ignoring numerical coefficients.

Using the continuity of the wave function on the NC surface, we get
\begin{equation}
\psi_0\simeq\frac{k_F}{\sqrt{d}}Y_l^m(\theta',\phi')\left\{ \begin{array}{ll}
\displaystyle{j_l(k_Fr')} &\quad r'<\frac{d}{2}\\
&\\
\displaystyle{\frac{j_l\left(k_Fd/2\right)}{h_l^{(1)}(id/2b)} h_l^{(1)}(ir'/b)} &\quad r'>\frac{d}{2},
\end{array} \right. \label{eq:21}
\end{equation}
where $h^{(1)}_l$ is the first-kind spherical Hankel function and only $Y_l^0(\theta',\phi')$ is nonzero at $\theta'=0$ corresponding to the state participating in the tunneling. The origin is set at the touching point of NCs with the $z$ axis pointed towards the center of the right NC and therefore the boundary plane is at $z=0$ (see the vertical dashed line in Fig. \ref{fig:3}). Again $(r',\theta',\phi')$ are the coordinates of $\boldsymbol{r'}=\boldsymbol{r}-\boldsymbol{r_L}$ in the spherical coordinate system and $\boldsymbol{r_L}$ is the coordinate of the center of the left NC.
For the finite potential barrier $U_0$ the derivative of the wave function divided by the effective mass is continuous across the surface, i.e.,
\begin{equation}
\left.\frac{d\psi_0}{dr'}\frac{1}{m^*}\right\vert_{r'-d/2=0^-}=\left.\frac{d\psi_0}{dr'}\frac{1}{m}\right\vert_{r'-d/2=0^+}.
\end{equation}
Using that $j_l(k_Fr')\simeq\sin\left[k_Fr'+\varphi_l\right]/k_Fr',\,h_l^{(1)}(ir'/b)\simeq be^{-r'/b}/r'$ at large $r'$ near the surface where $\varphi_l$ is a constant, we have
\begin{equation}
\begin{aligned}
\cot\left(\frac{k_Fd}{2}+\varphi_l\right)\simeq& -\frac{m^*}{mk_Fb}+\frac{2}{k_Fd}
\end{aligned}
\end{equation}
where $d/b\gg 1,\,m/m^*\gg 1$, $k_Fd\gg1$ at high doping concentration and $k_Fd\sim1$ for the ground state.
At $1/k_Fb\gg m/m^*$, the cotangent function diverges which means $\cos\left(k_Fd/2+\varphi_l\right)\approx 1,\,1/\sin(k_Fd/2+\varphi_l)\approx -m^*/mk_Fb$. So on the boundary plane inside the $b$-contact ($r'-d/2=0^+$), we have
\begin{equation}
\begin{aligned}
\psi_0\simeq&\frac{-k_F}{\sqrt{d}}\frac{mk_Fb}{k_Fdm^*}\sqrt{l}\\
\frac{d\psi_0}{dr}\simeq&\frac{k_F}{\sqrt{d}}\frac{mk_F}{k_Fdm^*}\sqrt{l}.
\end{aligned}
\end{equation}
The tunneling matrix element is then
\begin{equation}
\begin{aligned}
t\simeq\frac{\hbar^2k_F^3db^2(m/m^*)}{m^*d^2}.\label{eq:23}
\end{aligned}
\end{equation}
At $1/k_Fb\ll m/m^*$, the cotangent function either vanishes or is finite depending on whether $d/b\gg m/m^*$ or $k_Fd\gg 1$ is satisfied. This means the sine function is always finite and of the order 1. So inside the $b$-contact we get
\begin{equation}
\begin{aligned}
\label{eq:bigm}
\psi_0\simeq&\frac{k_F}{\sqrt{d}}\frac{1}{k_Fd}\sqrt{l}\\
\frac{d\psi_0}{dr}\simeq&\frac{-k_F}{\sqrt{d}}\frac{1}{k_Fdb}\sqrt{l},
\end{aligned}
\end{equation}
and the tunneling matrix element is
\begin{equation}
t\simeq\frac{\hbar^2k_Fd(m^*/m)}{m^*d^2}.\label{eq:24}
\end{equation}
One can check that when we put $k_F^{-1}\sim d$ into Eqs. \eqref{eq:23} and \eqref{eq:24}, we get the same tunneling matrix elements for the ground state as derived in Ref. \onlinecite{shabaev_dark_2013} for NCs touching in one point.

According to Eqs. \eqref{eq:23} and \eqref{eq:24}, the localization length is then
\begin{equation}
\xi\approx\left\{ \begin{array}{ll}
\displaystyle{\frac{d}{\ln\left[1/ndb^2(m/m^*)\right]}},&\quad\displaystyle{\frac{m}{m^*}\ll\frac{1}{n^{1/3}b}}\\
&\\
\displaystyle{\frac{d}{\ln\left[1/n^{1/3}d(m^*/m)\right]}},&\quad\displaystyle{\frac{m}{m^*}\gg\frac{1}{n^{1/3}b}}.
\end{array} \right. \label{eq:25}
\end{equation}
This leads to the critical concentration
\begin{equation}
n_c\simeq\left\{ \begin{array}{ll}
\displaystyle{\frac{1}{b^2d}\frac{m^*}{m}},&\quad \displaystyle{\frac{m}{m^*}\ll\left(\frac{d}{b}\right)^{1/2}}\\
&\\
\displaystyle{\frac{1}{d^3}\left(\frac{m}{m^*}\right)^3},&\quad\displaystyle{\frac{m}{m^*}\gg\left(\frac{d}{b}\right)^{1/2}},
\end{array} \right. \label{eq:26}
\end{equation}
which has its minimum value $n_c\simeq 1/(db)^{3/2}=1/\rho_b^3$ at $m/m^*\simeq\sqrt{d/b}$. Even this minimum value is much larger than $1/\rho_a^3$ since $\rho_a\gg\rho_b$. Thus, when NCs touch away from prominent facets the critical concentration is pushed much higher. In fact, for CdSe NC films, by using $b=0.1$ nm, $d=5$ nm, $m=m_e,\,m^*=0.13m_e$ \cite{Wheeler} where $m_e$ is the free electron mass, we get $n_c\simeq3\times10^{21}$cm$^{-3}$, which is difficult to achieve.

When NCs are separated by short ligands \cite{Murray_2005} by a small distance $s$, the overlapping wave functions exponentially decay as $\propto e^{-s/b}$ between neighboring NCs. Following a procedure similar to above derivations, we can get the tunneling matrix element $t$ as
\begin{equation}
t\simeq\frac{\hbar^2}{m^*d^2}\exp\left(-\frac{s}{b}\right)\left\{ \begin{array}{ll}
\displaystyle{k_F^3b^2d\frac{m}{m^*}},&\quad \displaystyle{\frac{m}{m^*}\ll\frac{1}{k_Fb}}\\
&\\
\displaystyle{k_Fd\frac{m^*}{m}},&\quad\displaystyle{\frac{m}{m^*}\gg\frac{1}{k_Fb}}.
\end{array} \right. \label{eq:sep}
\end{equation}
At smallest $k_F=1/d$, Eq. \eqref{eq:sep} gives the same results as in Ref. \onlinecite{shabaev_dark_2013} for NCs separated by short ligands.

Therefore, we get the localization length
\begin{equation}
\xi\approx\left\{ \begin{array}{ll}
\displaystyle{\frac{d}{s/b+\ln\left[1/ndb^2(m/m^*)\right]}},&\quad \displaystyle{\frac{m}{m^*}\ll\frac{1}{n^{1/3}b}}\\
&\\
\displaystyle{\frac{d}{s/b+\ln\left[1/n^{1/3}d(m^*/m)\right]}},&\quad\displaystyle{\frac{m}{m^*}\gg\frac{1}{n^{1/3}b}}.
\end{array} \right. \label{eq:22}
\end{equation}
At large $s$ we can ignore the logarithmic terms originating from the prefactor of $t$. But for small $s$, near the IMT, the role of these terms becomes important. One should note that even when NCs touch by short ligands, the localization length of electrons can be enhanced by increasing the doping concentration $n$ inside each NC.
The critical concentration $n_c$ is then
\begin{equation}
n_c\simeq\left\{ \begin{array}{ll}
\displaystyle{\frac{1}{b^2d}\frac{m^*}{m}\exp\left(\frac{s}{b}\right)},&\quad \displaystyle{s\ll b\ln\left[\left(\frac{m^*}{m}\right)^2\frac{d}{b}\right]}\\
&\\
\displaystyle{\frac{1}{d^3}\left(\frac{m}{m^*}\right)^3\exp\left(\frac{3s}{b}\right)},&\quad\displaystyle{s\gg b\ln\left[\left(\frac{m^*}{m}\right)^2\frac{d}{b}\right]}
\end{array} \right. \label{eq:27}
\end{equation}
which can easily become unrealistically large.

\section{Random-spectrum NC}
\label{sec:nond}

In previous sections, we have studied the highly degenerate case assuming that the splitting of $(n,l)$-shells is much smaller than the energy gap $\Delta$. In this section, we discuss limits of applicability of this assumption and find the localization length for strongly split and mixed $(n,l)-$shells which form a random spectrum similar to the case of metal garnules\cite{Beloborodov_2003,Ioselevich_2005,Beloborodov_2005,Glazman,Beloborodov}. We show below that this  happens at relatively small $\Delta/E_c < N^{1/3}$ or in the domain below the dashed line on Figs. \ref{fig:lowN} and \ref{fig:final2}. So the theory of this section is applicable for large enough NCs made from Si or ZnO.

Besides shifting the ladder of degenerate levels up and down discussed above, the random electric field created by neighboring charged NCs can split the degenerate shells of each NC due to the Stark effect. This field determined by nearest-neighbor NCs is ${\mathscr{E}}\sim e \sqrt{N} /\varepsilon_f d^2$. Electrons in the NCs respond to the internal field, which is smaller than ${\mathscr{E}}$ by the factor $3/(2+\varepsilon/\varepsilon_f)$. As we said in Sec. \ref {sec:sing}, $\varepsilon/\varepsilon_f$ is $ \simeq 3 $, so this factor is  $\simeq 3/5$ and we will ignore it.

To calculate the Stark splitting we first note that the matrix element of the electric field potential does not vanish only between shells with $l$ values differing by unity and is then of the order of $e{\mathscr{E}}d$. The typical energy difference between such shells in the spherical well with $N$ electrons is $N^{1/3}\Delta$. Therefore, the typical Stark energy shift or the width of the split shell $W$ emerges in the second-order perturbation theory and is
\begin{equation}
W\simeq\frac{(e{\mathscr{E}}d)^2}{N^{1/3}\Delta}.
\label{eq:Stark}
\end{equation}
(The Stark splitting can also come from random positions of $N$ donors inside each NC and is comparable to Eq. \eqref{eq:Stark}. This disorder creates an internal dipole moment $\sim \sqrt{N}ed$ and an electric field, oriented in a random direction.)

Comparing Eq. \eqref{eq:Stark} with the energy gap $\Delta$ between consecutive shells, we see that at
\begin{equation}
\Delta/E_c<N^{1/3}
\label{eq:10}
\end{equation}
the levels become random with the spacing $\delta=\Delta/N^{1/3}$ as the only characteristic energy \footnote{One may note that this criterion for degeneracy lifting is different from the one in Ref. \onlinecite{Ting}. However, since the critical concentration $n_c$ has the same expression in both degenerate and nondegenerate cases, this does not affect the correctness of the metal-insulator transition criterion obtained in Ref. \onlinecite{Ting}.}.
$\Delta/E_c=N^{1/3}$ is shown in Figs. \ref{fig:lowN} and \ref{fig:final2} by the dashed lines separating I and I' phases. When the inequality \eqref{eq:10} holds, the degeneracy is broken and different $(n,l)$-shells mix with each other. Thus inside each NC, the states close to the Fermi level and, therefore, involved in the electron tunneling have typically different $l$ numbers so that they have different parity and their tunneling matrix element $t$ has random signs. The electron wave functions of different $m$ hybridize and become chaotic instead of being confined in certain polar angles. So the typical magnitude of the wave function on the contact facet is $\sqrt{k_Fd}$ times smaller than that of the ``red" m=0 state for the degenerate case. These changes lead to the random matrix spectrum case which has been studied in previous work for larger dots\cite{Beloborodov_2003,Ioselevich_2005,Beloborodov_2005,Glazman,Beloborodov}.
In this case,
\begin{equation}
\begin{aligned}
\psi\simeq&\frac{k_F\rho}{d^{3/2}},\\
\frac{d\psi}{dz}\simeq&\frac{k_F}{d}
\end{aligned}
\end{equation}
where $k_F$ is given by Eq. \eqref{eq:Fermi_wave}.
Therefore, the typical tunneling matrix element is
\begin{equation}
\label{eq:28}
t\simeq\frac{\hbar^2k_F^2\rho^3}{m^*d^3}.
\end{equation}
At the same time, the energy gap between consecutive non-degenerate levels is also reduced to $\delta\simeq \Delta/(2l+1)\simeq \hbar^2/m^*d^3k_F$. Then according to Refs. \onlinecite{Ioselevich_2005,Beloborodov} the localization length is
\begin{equation}
\xi\approx\frac{d}{\ln\left(\sqrt{E_c\delta}/t\right)}.
\end{equation}
So one gets
\begin{equation}
\label{eq:deg}
\xi\approx\frac{d}{\ln\left(d/a_b^{1/2}n^{5/6}\rho^3\right)}
\end{equation}
where $a_b=\varepsilon_f \hbar^2/m^*e^2$ and $\varepsilon_f$ is the effective dielectric constant of the film.

According to Eq. \eqref{eq:deg}, at $t\simeq \delta$, the localization length is still much smaller than the NC diameter $d$, which seems to indicate a criterion different from $t\simeq\delta$ for the IMT. However, one should notice that as $t\rightarrow\delta$, the charge discreteness is no longer well preserved and the charging energy vanishes \cite{Matveev,Nazarov}, so $\delta$ takes the place of $E_c$ and changes the expression of $\xi$ to $d/\ln(\delta/t)$. Using Eq. \eqref{eq:28} and $\delta\simeq \hbar^2/m^*d^3k_F$, we get $t\simeq \delta$ at $k_F\rho\sim 1$. The localization length $\xi$ becomes comparable to $d$ at this point. This again leads to our above criterion Eq. \eqref{eq:result}, the same as for the degenerate case. Since this elimination of charging energy occurs in the vicinity of the IMT, we should see a steep growth of the localization length, which is a major feature different from the degenerate case. According to Eq. \eqref{eq:result}, $n_c\gg n_M$ at $\rho\ll a_B$. The critical concentration decreases with $\rho$ and saturates at $n_M$ when $\rho\sim a_B$.

\section{Conclusion}
\label{sec:con}

In this paper we studied theoretically what happens to the variable range hopping conductivity of semiconductor NC films when NCs are doped by donors with the concentration $n$. Experiments show that the localization length of electrons $\xi(n)$ grows with $n$ and at some $n=n_c$ becomes larger than the diameter $d$ of NCs, what signals that the film is approaching the insulator-metal transition (IMT). We provide theoretical estimates of $\xi(n)$ and $n_c$. The localization length is determined by the competition of disorder and transfer matrix element $t(n)$ between neighboring NCs.

We concentrated on the case of small spherical NCs in which the electron spectrum consists of degenerate energy shells separated by the quantization gap $\Delta$. In such films energy levels of NCs vary due to the dispersion of NC diameters and variations of the number of donors from NC to NC which result in random Coulomb potentials. We showed that for the standard diameter dispersion it is important for $\Delta/E_c > 5$, where $E_c$ is the charging energy, while the Coulomb disorder dominates for the opposite case $\Delta/E_c < 5$.

The matrix element $t(n)$ grows with $n$ and depends on the geometry of contacts between NCs. We calculated $t(n)$ for different types of contacts. We showed that for a finite separation between NCs  or even when NCs touch each other in one point, the IMT may need unrealistically large $n$. This is why we focused on the case when close-to-spherical NCs touch by smallest facets . We found $\xi(n)$ in this case and our results are in qualitative agreement with the experimental data for $\xi (n)$ obtained in Ref. \onlinecite{Ting}. For these facets $n_c$ is still relatively high and for $d=5$ nm CdSe NCs it corresponds to $N \sim 20$ electrons per NC, which justifies our large-$N$ approach. To make $n_c$ smaller one should deal with small NCs with $\Delta/E_c > 5$ and use NCs touching by larger facets. Another route is making much smaller dispersion of diameters, but this route does not look realistic.

$\phantom{}$
\vspace*{2ex} \par \noindent
{\em Acknowledgments.}

We are grateful to I. S. Beloborodov, A. S. Ioselevich, A. Kamenev, U. R. Kortshagen, B. Skinner, Al. L. Efros, and K. A. Matveev for helpful discussions. This work was supported primarily by the National Science Foundation through
the University of Minnesota MRSEC under Award No. DMR-1420013.

\bibliography{local}

\end{document}